\documentclass[12pt]{article}
\usepackage{graphicx}
\usepackage{amsfonts}
\usepackage{amssymb,amsmath,amscd}
\usepackage{color} 
\usepackage[top=2.75cm, bottom=2.75cm, left=2.75cm, right=2.75cm]{geometry}
\usepackage[pdftitle={Paper}, pdfauthor={Alejandro Cabrera}]{hyperref}

\newtheorem{proposition}{Proposition}

\newtheorem{lemma}[proposition]{Lemma}
\newtheorem{remark}[proposition]{Remark}
\newtheorem{definition}[proposition]{Definition}
\newtheorem{example}[proposition]{Example}

\def\+{{+\!\!\!+}}

\def\d{\partial}

\def\fF{\mathfrak{F}}
\def\fC{\mathfrak{C}}

\def\G{\Gamma}

\def\N{\nabla}

\def\h{\chi}

\renewcommand{\r}{{\rho}}
\def\l{\lambda}

\def\e{\varepsilon}

\def\pmb#1{\setbox0=\hbox{#1}%
\kern.0em\copy0\kern-\wd0
\kern-.04em\copy0\kern-\wd0
\kern.08em\copy0\kern-\wd0
\kern-.04em\raise.0433em\box0 }         %poor man's bold macro (TexBook)
\newcommand{\nc}{\newcommand}
\nc{\beq}{\begin{equation}}
\nc{\eeq}[1]{\label{#1}\end{equation}}
\nc{\ber}{\begin{eqnarray}}
\nc{\eer}[1]{\label{#1}\end{eqnarray}}
\nc{\pek}[1]{\cite{#1}}
\nc{\enr}[1]{(\ref{#1})}
\nc{\kal}[1]{{\cal{#1}}}
\nc{\dott}{\;\cdot\;}
\nc{\coker}{\mathrm{coker}}
\nc{\ie}{{\it i.e.}}
\nc{\eg}{{\it e.g.}}
\def\Z{{\mathbb Z}}
\def\R{{\mathbb R}}

\def\superx{{\boldsymbol x}}

\def\superxi{{\boldsymbol \xi}}
\def\superp{{\boldsymbol p}}

\def\0 {\nonumber}
\def\M{{\cal M}}
\def\N{{\cal N}}

\begin{document}
\setcounter{page}{0}
\newcommand{\inv}[1]{{#1}^{-1}} %inverse
\renewcommand{\theequation}{\thesection.\arabic{equation}}
\newcommand{\be}{\begin{equation}}
\newcommand{\ee}{\end{equation}}
\newcommand{\bea}{\begin{eqnarray}}
\newcommand{\eea}{\end{eqnarray}}
\newcommand{\re}[1]{(\ref{#1})}
\newcommand{\qv}{\quad ,}
\newcommand{\qp}{\quad .}

%%%%%%% Definizioni usate

\def\g{{\mathfrak{g}}}
\def\h{{\mathfrak{h}}}
\def\dv{{\rm div}}
\def\double{{\mathsf d}}

\def\x{{\underline{x}}}
\def\e{{\underline{\eta}}}
\def\l{{\underline{\lambda}}}
\def\r{{\underline{\rho}}}

\def\Seff{{S_{eff}}}

\def\ga{{ga}}

\def\fL{{\mathfrak{L} }}

\def\C{{\cal C}}

%\thispagestyle{empty}
%\begin{titlepage}
%\title{}
%\begin{flushright} \small
%UUITP-27/08  \\
%NORDITA-2008-65 \\
%\end{flushright}
%\smallskip

\thispagestyle{empty}
\begin{flushright} \small
UUITP-05/12
 \end{flushright}
\smallskip

\begin{center} \LARGE
{\bf AKSZ construction from reduction data}
 \\[12mm] \normalsize
{\large\bf Francesco Bonechi$^{a}$, Alejandro Cabrera$^{b}$ and Maxim Zabzine$^{c}$} \\[8mm]
{\small\it
$^a$I.N.F.N., Sezione di Firenze,\\
  Via G. Sansone 1, 50019 Sesto Fiorentino - Firenze, Italy \\
~\\
$^{b}$ Department of Mathematics, University of Toronto, Room 6290, 40 St.
George Street, Toronto, Ontario, Canada M5S 2E4 ;\\
 {\it Current Address}: Departamento de Matematica
Aplicada, Instituto de Matematica, Universidade Federal do Rio de
Janeiro, CEP 21941-909, Rio de Janeiro - RJ, Brazil. \\
~\\
$^c$Department of Physics and Astronomy,
Uppsala University, \\ Box 516, SE-751 20 Uppsala, Sweden
\\~\\
 
}
\end{center}
\vspace{10mm}
\centerline{\bfseries Abstract} 
We discuss a general procedure to encode the reduction of the
target space geometry into AKSZ sigma models.
This is done by considering the AKSZ construction with target the BFV model for constrained graded symplectic manifolds. 
We investigate the relation between this sigma model and the one with the reduced structure. We also discuss several examples in dimension two and three 
when the symmetries come from Lie group actions and systematically
recover models already proposed in the literature.
\bigskip

\eject

\tableofcontents

\begin{center}
{\normalsize {\small \textit{~}}}
\end{center}

%%%%%%%%%%%%%%%%%%%%%%%%%%%%%%%%%%%%%%%%%%%%%%%%%%%%%%%%%%%%%%%%%%%%%%%%%%%%%%%%%%%%%%%%%%%%%%%%%%%%%%%%%%%%%%%%%%%%%%%%%%%%%%%%%

\section{Introduction}
\label{Introduction}
The idea of encoding the reduction of geometric structures into sigma models is rather old and proved to be very useful. The main
motivation is to construct non trivial models with target $M/G$
by starting from the simpler one on $M$ and then \emph{gauging} it. The
geometrical reduction is then encoded in the gauge fixing of the new gauge degrees of freedom. 

In a series of recent papers \cite{Zuc07,Zuc08} the procedure of
gauging and reduction for the Poisson Sigma model (PSM in short) has been considered.
The model is constructed when there is an action of a Lie group $G$ on a Poisson manifold $M$ by Poisson diffeomorphisms and this action
is hamiltonian with momemntum map $\mu:M\rightarrow \g^*$. It is
important to note that this gauging
procedure is not standard since it takes into account the \emph{topological} nature of the PSM.

The main goal of this paper is to present a more general procedure which is
analogous to gauging but adapted to general topological sigma models
of AKSZ type as described below.

First, let us recall that the PSM is an example of the more general class of AKSZ (Aleksandrov-Kontsevich-Schwartz-Zaboronsky) topological field theories
\cite{Alexandrov:1995kv,Roytenberg:2006qz}. For any graded symplectic
\emph{finite dimensional} manifold $\M$ of degree $n$, which comes equipped with an
homological hamiltonian 
vector field, the AKSZ construction defines
a topological sigma model in dimension $n+1$. The manifold $\M$ is
called the \emph{target space} and the hamiltonian on it can encode in a
synthetic way various known geometrical structures. When the target manifold
is non-negatively graded, the case $n=1$ corresponds to Poisson
geometry and the corresponding model is the PSM; the case $n=2$
corresponds to Courant algebroid geometry and the model is called Courant sigma
model (CSM). 

Second, we recall that the graded manifold language offers a conceptually
simple framework to deal with the reduction of these
structures. Indeed, it becomes just ordinary \emph{symplectic reduction} in which the
reduced geometry is encoded in a \emph{reduced} graded symplectic manifold
$\M_{red}$ of the same type as the original $\M$. 
Poisson reduction as formulated for instance in \cite{L} is easily reformulated in the langauge of hamiltonian reduction of $n=1$
symplectic graded manifolds \cite{CZ}. The Courant algebroid reduction is the object of \cite{BCG} and its formulation in the graded language
can be found in \cite{BCMZ}.

Since one of the merits of the AKSZ method is the 
transparent relation between the sigma models and the underlying geometric structures, it seems reasonable
that the machinery developed for reducing the latter can be encoded in the AKSZ formalism
more efficiently and clearly than with the usual gauging procedures. In other words, it is natural to try to complete the
following diagram
\begin{eqnarray*}
  geometric \ structures & \overset{AKSZ}{\to} & sigma \ models \\
 \downarrow reduction & & \downarrow (?) \\
geometric \ structures & \overset{AKSZ}{\to} & sigma \ models
\end{eqnarray*}

In this paper we show that this can be done in a very general way by considering the so called BFV (Batalin-Fradkin-Vilkovisky)
construction. This emerged in \cite{BFV} as a hamiltonian version of BV-quantization
and it later evolved into the pure symplectic
geometry context (see \cite{Hen 85, Hen-Tei, Stasheff 96}).  In this paper, we shall use a variant of it
which takes a graded symplectic manifold together with reduction data
and produces another (higher dimensional) graded symplectic manifold $\M_{BFV}$ of the same type,
which we shall refer to as \emph{BFV manifold}. The BFV manifold encodes the reduction data
in the homological vector field and the key point is to use $\M_{BFV}$
as a target for the AKSZ construction. This enlarged topological field
theory is considered as the analogue of gauging
the original sigma model. 

Our main claim is that, by means of a formal argument, the AKSZ construction with
target $\M_{BFV}$ can be shown to compute the correlators of the AKSZ theory
corresponding to
the reduced target $\M_{red}$. The argument is just
formal since it needs to deal with integration over infinite
dimensional spaces of fields. In turn, we provide exact arguments in
the context of the
underlying \emph{zero modes} theory. We recall that the BV-space of zero 
modes is a finite dimensional theory which is a zero order
approximation of the full AKSZ model (see \cite{BZ 08,BMZ}) and which keeps
the overall structure as in the general AKSZ construction. It can thus
be taken as a partial verification of the full formal statement.

Finally, we present several examples in which the symmetries come from
Lie group actions. We show how our construction yields, as particular
cases, several
gauged models already considered
in the literature. We thus provide a more conceptual insight into
these models and clarify how to produce new ones in which other types of
reduced geometry can be also encoded.

\medskip

Let us now discuss the content of the paper in more detail.

In Section \ref{BVquantizationAKSZ} we review standard facts of $BV$
quantization and the AKSZ construction. 

In Section \ref{BFV manifold} we 
introduce the notion of {\it $BFV_n$ manifold}. We choose an 
axiomatic approach rather than the usual construction which is
postponed in Subsection \ref{BFVconstruction}. 
 This
construction is well known when applied to ungraded symplectic manifolds. Since we now
apply it to graded manifolds, we can incorporate an underlying
Q-hamiltonian into it. This results in the presence
of non-standard terms in the so-called BFV charge and we thus provide
details of its construction.

In Section \ref{BFV-AKSZ system} we study the AKSZ space of fields when the target is a BFV manifold. In this case the space of fields inherits an
additional grading and it is at the same time
a $\rm BFV_{-1}$ and a BV manifold. This is what we define to be a
{\it $BV-BFV$-manifold} and we show that it provides an homological model for BV-reduction.
 In
subsection \ref{BFV-BV}, we illustrate this by considering a toy model
which starts with $\fF=T^*[-1]M$, where the finite dimensional smooth manifold $M$ is
acted by a Lie group 
$G$. We show that the corresponding BV-BFV manifold $\fF_{BV-BFV}$ is defined by the choice of a volume form on the quotient $M/G$ and that the relevant
integrals computed on $\fF_{BV-BFV}$ coincide with those on the
quotient.
We describe in Subsection \ref{general_BFV_AKSZ} the general 
features of the AKSZ model with target a BFV manifold and we give a formal BV argument for the result shown in the finite dimensional example to be valid 
in general. 

The rest of the paper is devoted to discuss explicit examples. In Section \ref{groupactionn_1} we discuss
the case of $n=1$, {\it i.e.} the encoding of Poisson reduction in the Poisson Sigma Model. We show that in the case of a \emph{Poisson action} of Poisson-Lie 
group $G$, the ${\rm BFV}_1$ space is constructed by using the dual Poisson-Lie group $G^*$. Moreover, we extend the analysis
of the general toy model to the finite dimensional spaces of zero
modes. In Section \ref{AKSZ-BV-n_2} we discuss ${\rm BFV}_2$
spaces related to the reduction of exact Courant algebroids.

Finally, in Section \ref{conclusions} we present some conclusions and
further directions. The Appendices are used to examine particular cases and give details of some
arguments as indicated in the main text.

\medskip

During the elaboration of this work, the paper \cite{CMR12} came out
where a similar terminology is adopted. It is clear that it actually refers to a different construction. Indeed, in \cite{CMR12} 
the BV structure is defined on the space of fields on the bulk and the BFV on those on the boundary, while in our case they are defined
on the same space of fields (see also example \ref{ex: BFV_0}).

\bigskip
{\bf Notations and conventions}. In the literature there exist inequivalent definitions of a graded manifold. In this paper for a 
graded manifold $\M$ we always mean a supermanifold $\M$ endowed with a compatible $\Z$-grading, {\it i.e.} a coordinate atlas in which
each local coordinate $x^a$ is assigned a $\Z$-degree $\deg x^a$ and the coordinate transformations respect this degree. Moreover we will require that 
the parity is $\deg$ mod $2$. This grading is encoded by the {\it Euler vector field}, that in the local coordinates $\{x^a\}$ reads
$$
\epsilon = \sum_{a} (\deg x^a) x^a \frac{\partial}{\partial x^a}  \;.
$$

We denote with $C^k(\M)$ the global (polynomial) functions of degree $k$, {\it i.e.} $f\in C^k(\M)$ if $\epsilon(f)=k f$, and 
$C(\M)=\bigoplus_k C^k(\M)\subset C^\infty(\M)$. 
A symplectic form $\omega$ of degree $n$ is a closed, non degenerate 2-form such that $L_\epsilon \omega = n \omega$. It induces the structure of $n$-Poisson 
algebra on $C(\M)$. When $n$ is odd, a generator for the Poisson bracket is a degree $-n$ linear map $\Delta:C(\M)\rightarrow C(\M)$ satisfying
$$
\Delta(ab) = (\Delta a) b + (-)^{|a|}a (\Delta b) + (-)^{|a|}\{a,b\}~~~.
$$
We also require that $\Delta^2=0$ so that $\Delta$ becomes a derivation of the Poisson bracket. A ($-1$)-Poisson algebra together with a generator of the 
bracket will be called a {\it $BV$-algebra}.

Since in this paper there are a lot of different graded manifolds playing different roles, we try to help the reader by respecting the following notation:
we use roman letters $M$ for ordinary smooth manifolds, capital $\M$ for graded manifolds and $\fF$ for $BV$ manifolds.

\bigskip 
{\bf Acknowldegments}.
  A.C. wants to thank several institutions for
hospitality and support during the elaboration of
this project: HIM institute (Bonn) where he participated on the Trimester Program on Geometry and Physics
(2008) and both IMPA (Brazil) and the Math department of University of
Toronto where he was a post-doctoral fellow. A.C. also thanks H.-C. Herbig for fruitful discussions on the BFV construction. 
 F.B. wants to thank the authors of \cite{BCMZ} for sharing their preliminary draft, the Department of Physics and Astronomy at Uppsala University and the 
Nordita  for hospitality during the programs ``Geometrical Aspects of String Theory'' (2008) and ``Geometry of Strings and Fields'' (2011), when this project started and has been completed. The research of M.Z.  is supported by 
 VR grant  621-2011-5079.

\bigskip

%%%%%%%%%%%%%%%%%%%%%%%%%%%%%%%%%%%%%%%%%%%%%%%%%%%%%%%%%%%%%%%%%%%%%%%%%%%%%%%%%%%%%%%%%%%%%%%%%%%%%%%%%%%%%%%%%%%%%%%%%%%%%%%%%

\section{BV manifolds and AKSZ construction}\label{BVquantizationAKSZ}

We review in this section the notion of BV manifold and the AKSZ construction
of Topological Field Theories.

\subsection{BV quantization}
\label{BVmanifolds}
\begin{definition}
\label{def_BVmanifold}
A $BV$ manifold $(\fF,\omega_{-1},\nu)$ is a graded symplectic
manifold $\fF$, equipped with a symplectic form $\omega_{-1}$ of
degree $-1$ and a berezinian volume $\nu$ such that $C(\fF)$ is BV algebra. 
\end{definition}

Let $\Delta_\nu$ be the BV-laplacian induced by $\nu$ as 
$$
\Delta_\nu f = \frac{(-1)^{deg(f)}}{2} div_{\nu}X_f
$$
for $f \in C(\fF)$ and $X_f=\{f,\cdot \}$ its hamiltonian vector
field. The ring of global functions $C(\fF)$ is a (-1)-Poisson algebra and the BV laplacian $\Delta_\nu$ is a generator of the odd Poisson 
bracket satisfying $\Delta_\nu^2 =0$, so that it is a {\it $BV$-algebra}. 
A function 
$S\in C({\cal \fF})$, $\deg S=0$, solves the {\it Quantum Master Equation} (QME) if
$\Delta_\nu e^{\frac{i}{\hbar} S} = 0$. The QME can be equivalently written as
$$
\{S,S\}-2i\hbar \Delta_\nu S =0\;.
$$ 
The equation $\{S,S\}=0$ is called the {\it Classical Master Equation} (CME).  

%\noindent\textbf{Geometry of BV quantization.} \label{rmk: T*[-1]N} 
This is the geometric setup encoding the \emph{BV-quantization} \cite{BV1,BV2}, as formulated in \cite{Sch}. For every lagrangian submanifold 
${\cal L}\subset\fF$ the berezinian $\nu$ restricts to the berezinian $\nu_{{\cal L}}$ on ${\cal L}$. In \cite{Sch} it is shown that
for each $f\in C(\fF)$ such that $\Delta_\nu f=0$ and for two
cobordant lagrangian submanifolds ${\cal L},{\cal L}'\subset\fF$ one has
$$
\int\limits_{\cal L} f = \int\limits_{{\cal L}'} f     ~~~~~~~;
$$
while for each $g\in C(\fF)$
$$
\int\limits_{{\cal L}} \Delta_\nu g =0   \;.
$$

To briefly illustrate it, let us consider
a finite dimensional example. Let $N$ be a finite dimensional
smooth compact manifold and consider the cotangent bundle $\fF=T^*[-1]N$ equipped with the canonical symplectic form.
  This corresponds to
the canonical odd Poisson bracket $\{,\}$ of degree $+1$ (Schoutens bracket on multivectors on $M$); and homological charges are degree zero functions 
$S \in C(T^*[-1]N)$. The choice of a volume form
$\lambda$ on $N$ defines a berezinian $\nu_\lambda= \lambda \otimes \lambda\in\Gamma(\mathrm{Ber}(T^*[-1]N))=\Gamma(\det
(T^*N)\otimes\det (T^*N))$. For any submanifold $C\subset N$, $%
\sqrt{\nu_{\lambda}}:= \lambda|_C$ defines a berezinian for the lagrangian submanifold $%
N^*[-1]C \subset T^*[-1]N$, since $\mathrm{Ber} (N^*[-1]C)=\det (N^*C)\otimes
\det(T^*C)=$ $\det (T^*N|_C)$. The fundamental BV-theorems above are then a reformulation of the Stokes theorem on $N$
 (e.g., for a brief review see \cite{Qiu:2011qr}).

\subsection{The AKSZ construction}
\label{Section_AKSZ} 
The most interesting examples of $BV$ manifolds come as spaces of fields of Quantum Field Theory. In this case they are infinite dimensional spaces, 
so that the properties concerning integration must be implemented after renormalization. The AKSZ construction is a method introduced in 
\cite{Alexandrov:1995kv} that gives a solution of the classical master equation starting from very simple geometrical data.  
We describe it here following \cite{Roytenberg:2006qz}. 

Let us consider the following data:

\begin{itemize}
\item[$i$)] \it{The source}: {\rm A graded manifold $\N=T[1]N$, for
any smooth oriented manifold $N$ of dimension $n+1$, with $D=d$ being the de Rham
differential over $N$ and $\mu$ the canonical Berezinian measure
defined by orientation}.

\item[$ii$)] \it{The target}: {\rm A graded symplectic manifold
$(\M,\omega)$ with $\deg(\omega)=n$ and an homological vector field $Q$ preserving $\omega$%
. We require that $Q$ is Hamiltonian, \textit{i.e.} it exists $\Theta\in
C_{n+1}(\M)$ (functions of degree $n+1$) such that $Q=\{\Theta,-\}$.
Therefore $\Theta$ satisfies the following Maurer-Cartan equation}
\begin{equation*}
\{ \Theta, \Theta \}=0~.
\end{equation*}
\end{itemize}

\begin{remark}
In the above data, the source can be more general than a graded manifold; what is actually needed is a sheaf of finitely generated Frobenius algebras, see \cite{BMZ}
for details and the end of this subsection for the discussion of the AKSZ theory of zero modes.
\end{remark}

The space of maps $\fF=\mathrm{Map}(\N,\M)$ defines the space of
BV-fields for the theory (see below), where the BV structure
$(C(\fF),\{,\}_{\fF},\Delta)$ is given as follows.
First, if $%
\{u^\alpha,\theta^\alpha\}$ is a set of coordinates for $T[1]N$, then for any $f \in C(\M)$ we have
\begin{equation} \label{eq: ev*f}
 ev^*f=\sum_I f_{(I)}(u) \theta^I~,
\end{equation}
where $ev:\mathrm{Map}(\N,\M) \times \N \rightarrow \M$ denotes the evaluation map 
 \cite{CF01}, $\theta^I=\theta^{\alpha_1}...\theta^{\alpha_k}$ 
is a local basis for the fibers of $T[1]\N$. %and $f_{(I)}(u) \in C(\mathrm{Map}(\N,\M)) \otimes C(\N_0)$.
In particular, if $%
\{X^A\}$ are coordinates for $\M$, a superfield in $\fF$ is the collection $\Phi = \{\Phi^A\}$, where
\begin{equation*}
\Phi^A = \Phi^A_0(u) + \Phi^A_\alpha(u)\theta^\alpha +
\Phi^{A}_{\alpha_1\alpha_2}(u)\theta^{\alpha_1}\theta^{\alpha_2} + \ldots~.
\end{equation*}
The odd Poisson bracket
$\{,\}_{\fF}$ on $\fF$ comes from the symplectic form 
$$
\omega_{\fF}=\int\limits_\N \ ev^* \omega
$$
and the BV-laplacian is formaly given by $\Delta \simeq
\frac{\partial^2}{\partial \Phi_i \partial \Phi^i}$ for Darboux coordinates
  $\{X^A=x^i,p_i\}$ on the target $\M$. Notice that the laplacian is
  in general ill-defined and 
  needs appropriate regularization.

The AKSZ action reads as \beq
S[\Phi]= S_{kin}[\Phi]+ S_{int}[\Phi] = \int\limits_\N \mu ~\left (
\frac{1}{2} \Phi^A\omega_{AB} D\Phi^B + (-1)^{n+1} \Phi^*(\Theta)\right )~, %
\eeq{AKSZ-action} and solves the CME. We refer to this construction as
$AKSZ(\N,(\M,\omega,\Theta))$.

If the grading of $\cal M$ is non negative, then from \cite{Roytenberg:2006qz} we have a characterization of the encoded target geometry. For $n=1$ the
target manifold is $\M=T^{\ast }[1]M$, where $M$ is an ordinary smooth manifold and $\Theta =\pi \in C_{2}(\M)=\Gamma
(\Lambda ^{2}TM)$ is a Poisson structure. If we choose local coordinates $%
\{x^{i},b_{i}\}$ then $\Theta =\frac{1}{2}\pi ^{ij}b_{i}b_{j}$. The AKSZ
action defines what is known as \textit{Poisson Sigma Model} (PSM in short).
For $n=2$ then the target is $\M(E)$ the graded even symplectic manifold
associated to any vector bundle $E$ with nondegenerate symmetric bilinear
pairing and $\Theta $ is the hamiltonian associated to a Courant
algebroid structure
on it. Let us choose coordinates $\{x^i\}$ of degree $0$ on $M$ with
momenta $\{p_{x^i}\}$ of degree $2$, and a trivialization $\{e_A\}$ of
$E$ of orthonormal sections with coordinates $\{\lambda^A\}$ of degree
$1$. Denote also with $P_A^i$ and $T_{ABC}$ the coefficients of the
anchor and the bracket of the underlying Courant algebroid, 
respectively. We have that
\begin{equation}
\Theta =\lambda ^{A}P_{A}^{i}p_{x^i}-\frac{1}{6}T_{ABC}\lambda ^{A}\lambda
^{B}\lambda ^{C}~.  \label{hamiltonian_courant}
\end{equation}%
We will refer to the TFT defined by the AKSZ action as the \textit{Courant
Sigma Model}, (CSM in short).

%\noindent\textbf{The gauge fixing}. The correlators in the BV formalism are
%computed by choosing a lagrangian submanifold $\mathcal{L}\subset \fF$ of the space of
%fields and integrating observables over it. This
%is what we call gauge fixing. Given an observable $F \in C(\fF)$, the
%corresponding correlator is
%\begin{equation*}
%\langle F \rangle = \int_{\mathcal{L}} F e^{iS_{BV}} \;.
%\end{equation*}
%%%%%%%%%%%%%%%%%%%%%%%%%%%%%%%%%%%%

The choice of a Lagrangian
$\fL \subset \fF$ inside the space of BV-fields is called \emph{gauge
  fixing}. A quantum observable is an $F \in
C(\fF)$ satisfying $\Delta (F e^{iS_{BV}/\hbar})=0$. For such an $F \in C(\fF)$, one is interested in computing the integral expression
$$
\langle F \rangle := \int\limits_\fL F e^{\frac{i}{\hbar}S_{BV}} ~,
$$
which is called the (quantum) \emph{correlator}. As a consequence of the fundamental results of BV quantization, the 
correlators of quantum observables are independent of the gauge fixing
choice. 

%%%%%%%%%%%%%%%%%%%%%%%%%%%%%%%%%%%%

%\noindent\textbf{Boundary conditions}. If the source $\N_0$ has boundary then
%boundary conditions must be imposed. We require that superfields when
%restricted to the boundary take values in a lagrangian submanifold $\mathcal{%
%L\subset\M}$ such that $\Theta|_{\mathcal{L}}=0$. Moreover, if we express
%the kinetic term of the action in terms of a primitive $\theta$ of $\omega$,
%we realize that it actually depends on such a choice. In order to avoid it, we have to
%require that $\theta|_{\mathcal{L}}=0$.

%For $n=1$ boundary conditions are $\mathcal{L}=N^*[1]C$ for $C\subset M$
%coisotropic; for $n=2$ they correspond to the choice of a Dirac structure
%with support.

A finite dimensional analogue of this construction is the {\it $BV$ theory of zero modes}. It can be
seen as the symplectic reduction of the full theory with respect to the
constraint $D\Phi=0$. Indeed, this defines a coisotropic submanifold $\mathfrak{Z}$ of the space of
superfields $\fF$ such that the corresponding symplectic reduced space $\fF^Z=\fF // \mathfrak{Z}$ is finite dimensional and inherits all the
BV-structure from $\fF$. This reduced space $\fF^Z$ is {\it the
  BV-space of zero modes}. When $\partial N = \emptyset$, the reduced theory on the space of zero modes can be described again as an AKSZ construction where one
takes as source the de-Rham cohomology $H_{dR}(N)$ equipped with zero
differential. For further details see \cite{BZ 08,BMZ}.

Finally, we recall a useful concept of {\it effective action}, see \cite {losev} and \cite{Mn}. Let us consider a $BV$-manifold ${\cal F}$ that can be 
written as a product of $BV$-manifolds ${\cal F}_1\times{\cal F}_2$, 
such that $\Delta = \Delta_1+\Delta_2$. Let $S\in C({\cal
  F})$ be a solution of the quantum master equation (QME) $\Delta \exp i S/\hbar=0$. 
Let ${\cal L}_2$ be a Lagrangian submanifold of ${\cal F}_2$ and let us define the effective action $\Seff\in C({\cal F}_1)$ as
\begin{equation}\label{effective_action}
e^{\frac{i}{\hbar}\Seff}= \int\limits_{{\cal L}_2} e^{\frac{i}{\hbar} S}~.
\end{equation}
We clearly have that $\Seff$ solves QME on ${\cal F}_1$. Indeed we have that
$$
\Delta_1 e^{\frac{i}{\hbar} \Seff} = \int\limits_{{\cal L}_2} \Delta_1 e^{\frac{i}{\hbar} S} = \int\limits_{{\cal L}_2} \Delta e^{\frac{i}{\hbar} S} = 0~.
$$
We call ${\cal F}_1$ the space of infrared degrees of freedom and ${\cal F}_2$ the space of ultraviolet ones.
For instance if we take ${\cal F}$ as the space of superfields of the AKSZ construction and consider the splitting induced by the Hodge decomposition 
of $N$, where the infrared variables are the coefficient of the cohomology of $N$, we get that ${\cal F}_1=\fF^Z$ the $BV$-space of zero modes. 
Moreover the reduced BV-action is the lowest order of the expansion of
the effective action. When the target $\M$ is a general graded manifold and not just a 
vector space, then covariance with respect to the change of variables must be taken into account, see \cite{Bonechi:2011um}.

\bigskip \bigskip

%%%%%%%%%%%%%%%%%%%%%%%%%%%%%%%%%%%%%%%%%%%%%%%%%%%%%%%%%%%%%%%
 %%%%%%%%%%%%%%%%%%%%%%%%%%%%%%%%%%%%%%%%%%%%%%%%%%%%%%%%%%%%%%%

\section{The $BFV_n$ manifold}
\label{BFV manifold}

In this Section we introduce the BFV (Batalin-Fradkin-Vilkovisky) manifold. We give first an axiomatic definition modeled on the usual BFV 
construction of first class constraints. The main reason for this choice is to emphasize the ingredients that will be used in the AKSZ formulation, 
rather than all the details of the construction that will be sketched in Section \ref{BFVconstruction}.

\begin{definition}\label{def_BFV_space}
A $BFV_n$ manifold is a triple (${\cal M},\Omega_n,\Theta$), where ${\cal M}$ is a $\Z$-graded symplectic manifold with a degree $n$ symplectic 
form $\Omega_n$ and degree $n+1$ BFV-charge $\Theta\in C^{n+1}({\cal M})$ that satisfies $\{\Theta,\Theta\}=0$.  Moreover ${\cal M}$ is 
endowed with an extra $\Z$-grading $ga$ called the {\it ghost-antighost} degree, such that 
$\ga(\Omega_n)=0$ and $\Theta=\sum\limits_{k\leq 1} \Theta_k$, with $\ga(\Theta_k)=k$. 
\end{definition}

In the above definition we implicitly assumed that the $\Z$-degree $\deg$ 
determines the parity $\Z_2$-degree of ${\cal M}$ seen as a supermanifold. By having an extra $\Z$-grading we mean that there is an atlas of
coordinates homogeneous in the $ga$ degree and such that the changes of coordinates also respect this grading. Moreover we assumed the vanishing of the
coefficients $\Theta_r$ for $ga$ degree $r\geq 2$; all the examples we will
consider satisfy this property.

\begin{example} \label{ex: BFV_0}
Let $\M$ be a $\Z$-graded manifold with degree $0$ symplectic
form $\Omega_0$ and degree $1$ homological hamiltonian $\Theta \in
C^{1}({\cal M})$. It is easy to see that $(\M, \Omega_0, \Theta)$ is
an example of a $BFV_0$ manifold in the above sense if we take the
extra grading to be $ga=deg$. This encompasses the usual versions of the
BFV construction (see \cite{Kimura, Schaetz, Stasheff 96}). Also notice that some of these examples arise when $\M$ is
infinite dimensional, 
like the one obtained by applying the AKSZ construction to the
boundary $\d N$ of the source $N$ as done in \cite{CMR12}.
\end{example}

Let us denote with ${\cal M}_0$ and with ${\cal M}_{>}$ the submanifolds of ${\cal M}$ obtained by putting to zero the 
coordinates with non zero and negative $\ga$ degree, respectively. We have clearly ${\cal M}_0\subset {\cal M}_>\subset{\cal M}$. 
The restriction for the $BFV$-charge to have $ga$-degree less than $2$ is motivated by Lemma \ref{fromBFVtoReduction} and by the examples 
that we will discuss, see in particular Propositions \ref{prop: map from reduced cohomology} and \ref{BFV_poisson_lie}.

\smallskip
\begin{lemma}
${\cal M}_0\subset {\cal M}$ is a symplectic submanifold. 
\end{lemma}
{\it Proof}. Let $\Pi=\Omega_n^{-1}$ be the Poisson tensor. One immediately see that $\Pi|_{{\cal M}_0}$
does not mix coordinates with zero and non zero $ga$ degree, so that $\Omega|_{{\cal M}_0}$ is non degenerate. $\square$
\smallskip

Unlike the usual BFV-construction, in general, the ghost-antighost coordinates are not fibre coordinates of 
some vector bundle over the unconstrained ${\cal M}_0$,
{\it i.e.} non linear transformation rules are allowed, as in the next example.  

\begin{example}
The graded manifold description of standard Courant algebroid gives an example of $BFV_2$-manifold. Let ${\cal M}=T^*[2]T[1]M$, with Darboux coordinates 
$(x^\mu,b_\mu,\psi^\mu,p_\mu)$ with deg ($0,1,1,2$) and hamiltonian $\Theta=\psi^\mu p_\mu $.
The ga degree is defined as 
$$ga(x^\mu,b_\mu,\psi^\mu,p_\mu)=(0,-1,1,0)$$ 
so that $\Theta=\Theta_1$ and $\M_0 = T^*[2]M$. Notice that the nonlinear change
of coordinates in this case is $p_\mu \mapsto p_\mu + A_{\mu \lambda}
^{\nu}(x) b_\nu \psi^\lambda$, which indeed respects the extra $ga$-degree. 
\end{example}

We denote with $C^k({\cal M})$ the set of functions of $\deg=k$ and with $C^{k}_{p}({\cal M})$ 
those with $\ga = p$ and $\deg=k$. We thus write $Q=\sum\limits_{r\leq 1}Q_r$ the hamiltonian vector 
field, where $Q_r=\{\Theta_r,-\}: C^{k}_{p}({\cal M})\rightarrow C^{k+1}_{p+r}({\cal M})$, and with 
$H_{Q}$ its cohomology.

The local model for ${\cal M}$ reads as follows: there are coordinates in which
$\Omega_n=dx^\mu dp_\mu + d\xi^a dp_a$, where $\ga(x^\mu)=\ga(p_\mu)=0$ and 
$\ga(\xi^a)=-\ga(p_a)>0$. We call $\xi^a$ the ($BFV_n$) {\it ghosts} and $p_a$ the ($BFV_n$) {\it antighosts}. The submanifold ${\cal M}_0\subset {\cal M}$ 
is locally defined by $p_a=\xi^a=0$ and ${\cal M}_>$ by $p_a=0$.

Let us assume for simplicity that $\ga(\xi^a)=-\ga(p_a)=1$. The most general case of higher $ga$ degree ghosts and antighosts is related to
the reducibility of the constraints of the associated reduction data (see below) and these coordinates should be called called ghosts-for-ghosts.

We now analyze the content of the $BFV_n$-charge. If we decompose the equation
$\{\Theta,\Theta\}=0$ in the $\ga$ degree we get for each $r\leq 2$
$$
\sum_{k+\ell=r}\{\Theta_k,\Theta_\ell\} = 0\;.
$$
If we develop $\Theta_k$ on the ghosts/antighosts variables, {\it i.e.} $\Theta_0 = \theta + \theta^a_b \xi^b p_a+\ldots$, $\Theta_1 = \Theta_{1,a}\xi^a + 
1/2 \Theta_{1,ab}^c\xi^a\xi^b p_c+\ldots$, we get the following relations on ${\cal M}_0$
\begin{equation}\label{reduction_data_from_BFV}
\{\theta,\theta\}+2 (-)^{t_a} \Theta_{-1}^a\Theta_{1,a}=0~,~~~~\{ \theta , \Theta_{1,a}\} + (-)^{t_at_b}\theta^b_a \Theta_{1,b} = 0 \;, 
\end{equation}
$$
\{\Theta_{1,a},\Theta_{1,b}\} + \Theta_{1,c} \Theta^{c}_{1,ab} =0  \;,
$$
where $t_a=\deg \Theta_{1a}$.
%In the general case, since $Q$ doesn't preserve the $ga$ degree, $(C({\cal M}), Q)$ is just endowed of the decreasing filtration $F^pC(\cal A)$ 
%given by the $ga$ degree. The associated graded module in $ga$ degree $p$ is $C^{p,q}({\cal M})=F^pC^q({\cal M})/F^{p+1}C^q({\cal M})$. 
%The graded module associated to the filtration of the cohomology $H_{Q}$ is denoted as 
%$E^{p,q}_{Q}=F^p H^{p+q}_{Q}/F^{p+1} H^{p+q}_{Q}$. The first term of the spectral sequence is 
%$E_1^{p,q}=H^{p+q}(C^{p,\bullet},{Q}_0)$ with differential (induced by) 
%$Q_1: E_1^{p,q}\rightarrow E_1^{p+1,q}$. 
If $Q=Q_1$, as in the usual $BFV$-construction, then the complex $(C({\cal M}),Q)$ is bigraded by the $ga$-degree. 
%, $E_1^{p,q}=C^{p,q}({\cal M})$ and $E^{p,q}_{Q}=H^{p,p+q}_{Q}=H^{p+q}(E^{\bullet,q}_1, Q_1)$.

\begin{definition}
Degree $n$ 
{\it reduction data} are given by a quadruple ($\M_0,\omega_n,{\cal I},\theta$), where $\M_0$ is a graded manifold with symplectic form 
$\omega_n$ of degree $n$, ${\cal I}\subset C(\M_0)$ is a coisotropic ideal, {\it i.e.} $\{{\cal I},{\cal I}\}\subset {\cal I}$, and 
$\theta\in C^{n+1}(\M_0)$ is a function 
invariant under the coisotropic distribution and homological on the constraint, {\it i.e.} $\{\theta,{\cal I}\}\subset {\cal I}$ and 
$\{\theta,\theta\}\in {\cal I}$.
\end{definition}

The reduced algebra $(C^\infty(\M_0)/{\cal I})^{inv}$ is a $n$-Poisson algebra; the hamiltonian 
$\theta$ defines $\theta_{red}\in(C^\infty(\M_0)/{\cal I})^{inv}$ and let $Q_{red}=\{\theta_{red},-\}$ satisfying $Q_{red}^2=0$. Let us 
denote with $H_{Q_{red}}$ its cohomology. We call $f\in C(\M_0)$ a {\it reducible observable} if 
$\{\theta,f\}\in {\cal I}$ and $\{{\cal I},f\}\subset {\cal I}$. 
 
We shall say that the reduction data are {\it regular} if the ideal $\cal I$ defines a coisotropic submanifold ${\cal C}\subset \M_0$ and the 
quotient $q: {\cal C}\rightarrow \M_0//{\cal C}$ defines a smooth submersion. In this case, the reduction data induce a degree $n$ symplectic structure 
$\omega_{red}$ and an hamiltonian homological vector field $Q_{red}$ on the
reduced graded manifold $ \M_0//{\cal C}$. In the regular case we denote the reduction data also with ($\M_0,\omega_n,{\cal C},\theta$).

%In this case, such an $f$ induces a $Q_{red}$-closed $f_{red}\in C(\M//{\cal C})$ satisfying $f|_{\cal C}=q^*f_{red}$. 

\begin{lemma}\label{fromBFVtoReduction}
A $BFV_n$ manifold (${\cal M},\Omega_n,\Theta$) defines the degree $n$ reduction data
$({\cal M}_0,\Omega_n|_{{\cal M}_0},$ ${\cal I}_{\Theta_1},\theta=\Theta|_{{\cal M}_0})$, where ${\cal I}_{\Theta_1}$ is locally generated by $\Theta_{1,a}$. 
\end{lemma}
{\it Proof.} First, notice that $\Theta_{1,a}=\frac{\partial}{\partial
  \xi^a} \Theta_1 |_{\M_{>}}$ transforms covariantly under change of
coordinates and thus ${\cal I}_{\Theta_1}$ is globaly well defined. The equations (\ref{reduction_data_from_BFV}) 
imply that ${\cal I}_{\Theta_1}$ is coisotropic and that $\theta$ is an invariant homological 
function. $\square$

\medskip

Given the reduction data ($\M_0,\omega_n,{\cal I},\theta$) and a ${\rm BFV}_n$ manifold
(${\cal M},\Omega_n,\Theta$) inducing them, we will say that ${\cal M}$
is a \emph{$BFV_n$-model} for the reduction data. $BFV_n$-models are in general not unique, see for example
\cite{Schaetz} for the $n=0$ case.

\medskip

Let $F=\sum\limits_{r\leq 0} F_r \in C({\cal M})$ be concentrated in 
nonpositive $ga$-degree and $Q$-closed. It is easy to check that $f=F|_{{\cal M}_0}$ is a reducible observable. 

\begin{remark}\label{obstructiontocohomology}
In the general case, the correspondence $F\rightarrow F|_{{\cal M}_0}$ does 
not define a map in cohomology. Let us consider $F+Q(G)$ with $Q(G)_r=0$ for $r>0$. Let us denote 
$G_0=g+\ldots$, $G_1=g_a \xi^a +\ldots$, $G_{-1}=g^ap_a+\ldots$. One computes
$$ Q(G)|_{{\cal M}_0}=\{\theta,g\} + (-)^{|g|+(\lambda_a+1)(n+1)}g^a \Theta_{1,a} + g_a \Theta^a_{-1} (-)^{(|g|+1)(n+1)+\lambda_a n)}  ~~~~.$$
Even if $\Theta^a_{-1}=0$, as it happens in Proposition \ref{BFV_poisson_lie}, $g=G|_{{\cal M}_0}$ doesn't 
automatically define a primitive for $(Q(G))_{red}$. Indeed, since $Q(G)_1=Q_1 G_0+Q_0 G_1+ Q_{-1} G_2+\ldots=0$, 
we have in lowest degree in $\xi$
\begin{eqnarray}\label{QG1=0}
0&=& \{\theta, g_a\} +\{\Theta_{1,a},g\} (-)^{(n+1+\lambda_a)(|g|-n)}+ \Theta_{1,b} g^b_a (-)^{(1+\lambda_b)(|g|+1)} \cr
&+& \theta_a^b g_b (-)^{\lambda_a\lambda_b +(\lambda_a+\lambda_b)(|g|+n+1)}
+ \Theta_{-1}^b g_{ba} (-)^{(1+\lambda_b)(|g|+1)} ~~~~,
\end{eqnarray}
so that $g$ need not be invariant and does not define a $Q_{red}$-primitive for $(Q(G))_{red}$.
\end{remark}

\medskip
We recall that if $Q=Q_1$ then $H_Q=\sum\limits_{p,q}H_Q^{q,p}$ is bigraded and let us denote $H_Q^{ga=0}={\mathop \oplus \limits_{p}}H^{0,p}_Q$.
\smallskip 

\begin{proposition}
\label{cohomology_nonpositive_gadegree}
If $Q=Q_1$, then the map $F\rightarrow F|_{{\cal M}_0}$ for $Q$-closed
$F=\sum_{r\leq 0} F_r$, concentrated in non positive $ga$-degree,
descends to a map
$$
\Psi: H^{ga=0}_Q \rightarrow (C(\M_0)/{\cal I})^{inv}\;.
$$ 
\end{proposition}

{\it Proof}.  With the same notation of Remark \ref{obstructiontocohomology}, the result follows from (\ref{QG1=0}) observing that $g$ is 
invariant and so it reduces to a primitive for
$(Q(G))_{red}$. $\square$

More generaly, in the above case, one can compute the cohomology $H_Q$
using spectral sequence arguments (see \cite{Stasheff 96} for the $n=0$ case).

%\setcounter{subsection}{-1}

%%%%%%%%%%%%%%%%%%%%%%%%%%%%%%%%%%%%%%%%%%%%%%%%%%%%%%%%%%%%%%%%%%%%%%%%%%%%%%%%%%%%%%%%%%%%%%%%%%%%%%%%%%%%%%%%%%%%%%%%%%%%%%%%%%%%%%%%%%%%%%%%%%%%%%%

\subsection{The $BFV_n$ construction from reduction data}
\label{BFVconstruction}

We sketch in this subsection the construction of a $BFV_n$-model from the regular reduction data ($\M_0,\omega_n, 
\C,\theta)$. The case $n=0$ and $\theta=0$ is standard, see \cite{CH} for more 
details on the generic case (see also \cite{LS} for
the $n=1$ case). 

In general, this $BFV_n$ construction will also depend on non-canonical choices
such as a tubular neighbourhood and a connection on the normal bundle
to $\C$, as for $n=0$ (see for instance \cite{Schaetz}). In this paper, 
however, we will only consider cases in which $\C$ is given as zero level set of a map
$\mu:\M_0 \rightarrow W$ where $W$ denotes a graded vector space. We shall also make the regularity assumption\footnote{In the non-regular case, one needs to introduce
ghosts-for-ghosts of higher $ga$-degree in order to kill unwanted
cohomology (see e.g. \cite{Stasheff 96}). The construction goes along similar lines.}  that $0 \in W$ is a regular value for $\mu$, so that
($\M_0,\omega_n,\mu^{-1}(0),\theta$) define a set of regular reduction data. Our $BFV_n$ construction thus depends explicitly on the map $\mu$
 chosen.

After choosing a basis in $W$, the components of the moment map $\mu_a$ satisfy
\begin{equation}\label{eq: reducible theta}
\{\theta,\theta\}=\theta^a \mu_a~,~~~~
\{\mu_a,\mu_b\}_{\omega
}=\mu_{c}F_{ab}^{c} ~,~~~~~~~
\{\theta ,\mu_a\}_{\omega }
=\theta_{a}^{c}\mu_{c}  ~~~~.
\end{equation}
We set the graded manifold $\mathcal{M}=M_{n}\times T^{\ast }[n]V$
where $V=W^*[n+1]$. The Darboux coordinates on $T^*[n]V$ are $(\xi^a,p_a)$, called {\it ghosts} and {\it antighosts} respectively,
with degrees assignements  $\deg(\xi^a)=n+1-\deg(\mu_a)$ and $\deg(p_a)=\deg(\mu_a)-1$ and $\ga(\xi)=-\ga(p)=1$. The symplectic structure is  
$\Omega_n=\omega_n \oplus \varpi_n$, where $\varpi_n=d\xi^adp_a$. 

The reduction data fixe the lower terms of the BFV-charge
\begin{equation} \label{eq: struc tilde theta}
\Theta_0 = \theta + \theta_{a}^{c}\xi^ap_{c}
+O^{2}(p)
\end{equation}
\begin{equation} \label{eq: general BFV charge not modified}
\Theta_1 =\mu_{a}\xi^{a}+ (\pm)\frac{1}{2}F_{ab}^{c}\xi^{a}\xi^{b}p_{c} +O^{2}(p)~~,
\end{equation}%
so that the equation $\{\Theta,\Theta\}=0$ is satisfied up to second order in the antighosts. The sign in (\ref{eq: general BFV charge not modified}) 
can be fixed by comparing it with (\ref{reduction_data_from_BFV}).

A full solution for the BFV-charge $\Theta$ can be obtained by using the so called homological
perturbation method (\cite{Stasheff 96},\cite{Kimura},\cite{Herbig})
as follows. 
Let $res$ be the polynomial degree of the antighosts $p_a$ and let $\delta=\mu_a\frac{\partial}{\partial p_a}$. By degree considerations 
it is clear that $\delta^2=0$; let us denote with $H_\delta$ its cohomology. We denote with $F^{(r)}$ the components (of functions and vector fields) with
$res=r$. Remark that this additional grading is not defined in general for the $BFV_n$ space of Definition \ref{def_BFV_space}.

As a consequence of regularity, the homology of $\delta$ is concentrated in degree $0$, {\it i.e.} $H_\delta^r=0$ for $r>0$ (see Theorem 9.1 in 
\cite{Hen-Tei}).

\smallskip
The proof of the following theorem is standard for the case 
$\theta=0$ and $n=0$ (see for instance \cite{Hen-Tei}). The general case is similar but observe that  
terms with negative $ga$ grading appear in $\Theta$ due to the presence of nonvanishing $\theta$.

\smallskip
%%%%%%%%%%%%%%%%%%%%%%%%
\begin{proposition} \label{prop: map from reduced cohomology}
%Let the reduction data ($M_n,C_n,\omega_n,\theta$) be regular. 
\begin{itemize}
\item[$i$)]
There exists an homological function $\Theta=\sum\limits_{r\leq 1}\Theta_r$ coinciding with 
(\ref{eq: general BFV charge not modified}) up to first order in the antighosts.
\item[$ii$)] For any reducible observable $f\in C(\M_0)$ there exists $Q$-closed $F^\infty=\sum\limits_r F^{(r)}\in C({\cal M})$, where
$F^{(0)}=f$,
$res\ F^{(r)}=r$ and $F^{(r)}=\sum\limits_{l\leq 0} F^{(r)}_l$, where $ga(F^{(r)}_l)=l$. This correspondence defines a map 
\begin{equation}\label{eq: general ag=0 BFV cohomology} \lambda:H_{Q_{red}} \rightarrow H_{Q} ~~.
\end{equation}
\item[$iii$)] If $\theta=0$ then $Q=Q_1$ and $\lambda$ inverts the map
  $\Psi$ defined in Proposition \ref{cohomology_nonpositive_gadegree}.
\end{itemize}
\end{proposition}
%%%%%%%%%%%%%%%%%%%%%%%

{\it Proof}.

$i$) Let $\Theta^{(0)}=\theta+\mu_a\xi^a=\Theta^{(0)}_0+\Theta^{(0)}_{1}$ and using (\ref{eq: reducible theta}) let us compute
$$
\{\Theta^{(0)},\Theta^{(0)}\} = -2 \delta \Theta^{(1)}~~~,
$$ 
where $\Theta^{(1)}=\sum\limits_{r=-1}^1\Theta^{(1)}_r$. Define $\Theta^{\leq 1}=\Theta^{(0)}+\Theta^{(1)}$; 
by direct computation we get $res(\{\Theta^{\leq 1},\Theta^{\leq 1}\})\geq 1$. 
This procedure can be iterated. Let $\Theta^{\leq k}=\sum\limits_{l=0}^k\Theta^{(k)}$,
with $\Theta^{(l)}=\sum\limits_{r\leq 1}\Theta^{(l)}_r$, satisfy $res(\{\Theta^{\leq k},\Theta^{\leq k}\})\geq k$ and we can thus write
$\{\Theta^{\leq k},\Theta^{\leq k}\}=r^{(k)}+(res \geq k+1)$. Since by Jacobi identity
$$
0=\{\Theta^{\leq k},\{\Theta^{\leq k},\Theta^{\leq k}\}\} =  \delta r^{(k)} + (res \geq k)
$$
we get that $\delta r^{(k)}=0$, so that by regularity assumption $r^{(k)}=-2 \delta\Theta^{(k+1)}$. By taking into account that $ga(\delta)=1$ we also easily see
that we can choose $\Theta^{(k+1)}$ such that $\Theta^{(k+1)}=\sum\limits_{r\leq 1} \Theta^{(k+1)}_r$. Then $\Theta^{\leq k+1}=\sum\limits_{l=0}^{k+1} \Theta^{(l)}$
satisfies $res(\{\Theta^{\leq k+1},\Theta^{\leq k+1}\}) \geq k+1$. We can then get $\Theta=\sum_{k\geq 0} \Theta^{(k)}$ by induction.

$ii$) %Denote by $I_{C_n}=<\mu_a>$ the vanishing ideal of $C_n \subset M_n$ and $\tilde{I}_{C_n}=<\mu_a,p_{a}>$ the vanishing ideal of $C_n \times V \subset \mathcal{M} \simeq M_n \times T^*[n]V$.
The BFV charge (cf. eqs. 
$( \ref{eq: general BFV charge not modified} , \ref{eq: struc tilde theta} )$) is decomposed in the $res$ degree as
\begin{equation*}
 Q=\delta + \sum_{l\geq 0} Q^{(l)} ~~~.
\end{equation*}  
Since $f$ is reducible, we have $\{\theta,f\}=f^a\mu_a$ and $\{\mu_a,f\}=f_a^b\mu_b$, so that one computes
\begin{equation*}
Qf=\{ \Theta, f \} = -\delta F^{(1)} + (res \geq 1)\;,
\end{equation*}
where $F^{(1)}= F^{(1)}_0+F^{(1)}_{-1}$. Define $F^{(\leq 1)}=f + F^{(1)}$ and compute
$QF^{(\leq 1)}=(res \geq 1)= r^{(1)}+(res \geq 2)$ where $r^{(1)}$ denotes the $res=1$ part of $QF^{(\leq 1)}$. 
Thus, from $Q^2=0$ we get $ 0=Q^2F^{(\leq 1)}=\delta(r^{(1)}) + (res \geq 1)$ so that $\delta(r^{(1)})=0$. By the regularity assumption, there must 
exist $F^{(2)}$ such that $\delta F^{(2)} = -r^{(1)}$ and $resF^{(2)}=2$. Moreover, since the non vanishing
terms of $r^{(1)}$ appear with $ga\leq 1$, then $F^{(2)}$ can be chosen with nonvanishing $ga$-components 
only for $ga\leq 0$. The procedure can be iterated and adding corrections analogously, we get the representative 
$F^{\infty}(f)=\sum_i F^{(i)}$. 

%\begin{eqnarray}
% \nonumber QF^{\infty}=0 
%\label{eq: elements on image of map}
%\end{eqnarray}

Let now $f_{red}$ be $Q_{red}$ exact, {\it i.e.} let $f=\{\theta,g\}+\mu_a g^a$ and $\{\mu_a,g\}=g_a^b \mu_b$. We want to find $G$ such that 
$F^\infty(f)=Q(G)$. The components in $res$ degree must satisfy for each $k\geq 0$
\begin{equation}\label{primitive_resdegree}
\delta G^{(k)} = F^{(k-1)} -\sum_{r=0}^{k-1} Q^{(k-1-r)}(G^{(r)})\;.
\end{equation}
One can check that with $G^{(0)}=g$ and $G^{(1)}= p_a(g^a \pm g^a_b\xi^b)$ it is satisfied with $k=0,1$ (the sign can be easily fixed, but 
it is irrelevant here). We proceed by induction and 
let us assume that it exists $G^{(l)}$, $l\leq k$, satisfying (\ref{primitive_resdegree}) for $k\geq 1$. In order to show that it is true for $k+1$, by the regularity of the constraint 
it is enough to show that 
$$
\delta\left(F^{(k)}-\sum_{r=0}^k Q^{(k-r)}(G^{(r)})\right)=0\;,
$$
which is done by a straightforward computation. Also, the ambiguity involved in 
choosing $\delta$ potentials translates into $Q$ exact terms, so that we get in cohomology the map $(\ref{eq: general ag=0 BFV cohomology})$ we wanted. 

$iii$) If $\theta=0$ then it is clear from $i$) that $\Theta^{(k)}=\Theta^{(k)}_1$ and from $ii$) that $ga(F^\infty(f))=0$. Moreover, 
$\Psi\circ\lambda={\rm id}$ is obvious, and $\lambda\circ\Psi={\rm id}$ can be proven similarly. $\square$

\medskip
\medskip

\begin{remark}
If $\Theta=\Theta_1$, as in the usual $BFV_0$ construction, then the $BFV$ cohomology is graded by $ga$-degree and the image of $\Psi^{-1}$ described in Proposition 
\ref{prop: map from reduced cohomology} is $H^{ga=0}_Q$. As a
corollary we get that, in this case, $H^{ga=0}_Q=C(\M_0//\C)$ as graded Poisson
algebras.
\end{remark}

%\begin{remark} \label{rmk: BFV_1=T*[1]BFV_0}
% Let $\mu^{(0)}=J:M_0 \rightarrow V_0$ be a map such that $C_0=J^{-1}(0)$ is coisotropic inside $(M_0,\pi)$. Denote by $\mu^{(1)}_a=v_a=\{J_a, \ \}$ the corresponding hamiltonian vector fields and assume the regularity condition that $Span\{v_a \} \subset TM_0$ has constant rank. Setting $C_1=\{\mu^{(0)}_a=0,\mu^{(1)}_b=0 \} \subset T^*[1]M_0$ one then has
%\begin{equation*}
% BFV_1(T^*[1]M_0,\omega_{can},C_1)= T^*[1]^\theta(BFV_0(M_0,\pi,C_0))
%\end{equation*}
%where $ T^*[1]^\theta (N,Q)$, for $(N,Q=Q^i \partial_{x^i})$ being a Q-manifold, denotes the space $T^*[1]N$ together with the degree $2$ function $\theta_Q=Q^i p_{x^i}$ which is the cotangent lift of $Q$ to $T^*[1]N$. See also remark \ref{rmk: moment map reduction} below for details on the case of moment maps.
%\end{remark}

\begin{remark}\label{vertical_complex}
There is another useful object related to BFV cohomology for $ga\geq 0$, namely its \emph{vertical complex} (cf \cite{Herbig}).
It is given by $\mathcal{V}:=(C(\mathcal{M}_>|_{\C}),d_v)$ where $d_v \circ r=r \circ Q$,
being $r:C(\mathcal{M}) \to C(\mathcal{M}_>|_{C_n})$ the
restriction to the submanifold $\mathcal{M}_>|_{C_n} \subset
\mathcal{M}$ defined by setting all negative $ga$ coordinates to zero
and by restriction to the coisotropic $\C \subset \M_0$ for the $ga$ degree
zero coordinates. One can use arguments similar to the above proof to show (\cite{CH}) that the 
map $r$ induces an isomorphism from the BFV cohomology of $(C^{\cdot }(\mathcal{M}),Q)$ 
to the cohomology of $\mathcal{V}$. There is also a map $i: C(\C) \to
\mathcal{V}$ given by the inclusion of the $ga$ degree zero part and one
has that the map $(\ref{eq: general ag=0 BFV cohomology})$ factors as
$\lambda = r^{-1}\circ i\circ q^*$. It is, in general, the map induced
by $i \circ q^*$ in cohomology the one that might be non injective.
\end{remark}

%%%%%%%%%%%%%%%%%%%%%%%%%%%%%%%%%%%%%%%%%%%%%%%%%%%%%%%%%%%%%%%%%%%%%%%%%%%%%%%%%%%%%%%%%%%%%%%%%%%%%%%%%%%%%%%%%%%%%%%%%%%%%%%%%%%%%%%%%%%%%%%%%%%%%%%%%%%%%%%%

\section{The AKSZ-BFV system}
\label{BFV-AKSZ system}
%%%%%%%%%%%%%%%%%%%%%%%%%%%%%%%%%%%%%%%%%%%%%%%%%%%%%%%%%%%%%%%%%%%%%%%%%%%%%%%%%%%%%%%%%%%%%%%%%%%%%%%%%%%%%%%%%%%%%%%%%%%%%%%%%%%%%%%%%%%%%%%%%%%%%%%

We consider here a field theoretical realization of the homological reduction described in the previous sections. We will introduce first spaces
that are at the same time $BV$ and $BFV_{-1}$. Indeed the case $n=-1$ is special in the $BFV$ construction, since the $BFV_{-1}$ space is a symplectic 
manifold of degree $-1$ endowed with a charge of degree \emph{zero}. The homological condition can be seen as the CME and it is natural to require 
the existence of a berezinian such that
the charge solves the QME too. This is what we will call a {\it BV-BFV manifold}. We show that it must be seen as an homological model for the reduction of $BV$-manifolds.
In subsection \ref{BFV-BV} we discuss the reduction of the most basic example of BV-manifold, $T^*[-1]N$, with respect to a group action. 
In this example we understand concretely how the structure of BV-BFV manifold encodes the reduction of the BV structure.
In subsection \ref{general_BFV_AKSZ} we discuss the case we are most interested in, {\it i.e.} the AKSZ construction 
in dimension $d+1$ having a $BFV_d$ manifold as target.

\subsection{BV-BFV manifolds}
\label{BFV-BV}

\begin{definition}\label{def_BV-BFV}
A $BV-BFV$ manifold is a $BFV_{-1}$ manifold $(\fF,\omega_{-1},\Theta)$ endowed with a berezinian $\nu$ (so that it becomes a BV-manifold) such 
that $\Theta$ solves the QME and $ga(\Delta_\nu)=0$. 
\end{definition}

This definition implies that on the $BFV$ charge we impose $\Delta_{\nu} \Theta=0$; 
as a consequence, if $Q=\{\Theta,-\}$, then  $[Q,\Delta_\nu]=Q\Delta_\nu+\Delta_\nu Q=0$ and the cohomology $H_Q$ of the $BFV$ differential 
inherits a generator for its odd Poisson bracket, 
{\it i.e.} it is a $BV$-algebra. Since $ga(\Delta)=0$ then $\fF_0\subset\fF$, obtained by putting 
all ghosts and antighosts to zero, is a $BV$ manifold (or better $C(\fF_0)$ is a $BV$-algebra). 

This definition gives an insight into the reduction of $BV$-manifolds.
Let $\Theta=\Theta_1$ be concentrated
in $ga$ degree $1$ so that the restriction $F\rightarrow F|_{\fF_0}$ descends to a map $\Psi: H^{ga=0}_Q\rightarrow C(\fF_0//C_{\Theta_1})$, 
as in Proposition \ref{cohomology_nonpositive_gadegree}. If the constraints defined by $\Theta$ are regular then $\Psi$ is an isomorphism and 
$\Delta_\nu$ induces a generator of the Poisson bracket on $C(\fF_0//C_{\Theta_1})$. 
We want to stress here the role of the regularity condition on the constraints. Indeed the same regularity condition that allows the
BFV construction out of reduction data allows the reduction of the BV laplacian.
Let us spell out the conditions given by Definition \ref{def_BV-BFV}. By using the same notations as in Section \ref{BFVconstruction}, we see that 
the QME equation for $\Theta$ implies $\Delta_\nu \mu_a + F_{ab}^b=0$. Let $f\in C(\fF_0)$ be reducible, {\it i.e.} $\{\mu_a,f\}=f_a^b\mu_b$; if $f$ is in the image of
$\Psi$ then the induced $BV$-generator is 
$$\Delta_\nu'f= \Delta_\nu f + f_a^a~~~~.$$ 
For a generic reducible $f$ this definition 
depends on the choice of $f_a^a$. Let $g^b_a$ satisfying $g_a^b\mu_b=f_a^b\mu_b$, then $\delta[(g_a^b-f_a^b)p_b\epsilon_{ab}]=0$, for $\epsilon_{ab}=\pm 1$, 
and by the regularity condition we get that $f_a^b-g_a^b= c_a^{bc}\mu_c$. Moreover, one computes that 
$$
\{\mu_a,\Delta_\nu'f\} = (-)^{1+t_a} (\Delta_\nu f_a^b)\mu_b + j_a(f)
$$
for some $j_a(f)$ to be computed explicitly. One checks that the regularity assumption implies that $j_a(f)=j_a^b \mu_b$
so that $\Delta_\nu'$ defines a generator for the Poisson bracket of $C(\fF_0//C_{\Theta_1})$.

\smallskip
Let us now discuss how to get a $BV-BFV$ manifold starting from the finite dimensional model $\fF_0:=T^*[-1]N$ for the
 space of BV-fields considered in Section
 \ref{BVmanifolds} when $N$ is acted by a Lie group $G$. Let $v_a \in \mathcal{X}(N)$
 be the vector fields generating the $G$-action on $N$.
This action can be lifted to
 $T^*[-1]N$ where it becomes hamiltonian with moment map components given by the
 degree $-1$
 functions $C(T^*[-1]N)\ni \mu_a \cong V_a$. Then, $\fC_\mu := \{\mu_a =0
 \} \subset T^*[-1]N$ defines a coisotropic submanifold. If the
 quotient map $q:N \to N/G$ is a smooth submersion, the coisotropic reduction yields 
$$
\mathfrak{F}_{red}: =T^*[-1]N // \fC_\mu \cong T^*[-1](N/G)~.
$$
  
Now, let $S_0 \in C(T^*[-1]N)$ be a degree zero,
  $G-$invariant ($\{ \mu_a,S_0  \} =0$)
  homological charge, so that $S_0$ reduces to $S_{red} \in C(\fF_{red})=C(T^*[-1](N/G))$. In this case,
  the $BFV_{-1}$ construction applied to the regular reduction data
  $(\fF_0=T^*[-1]N,\omega_0, \fC_\mu,S_0)$ yields  
  \begin{eqnarray*}
(\fF &:=&T^*[-1](N \times \g[1]), 
  \omega_{can}, S) \\
    S&=& S_0 + \Theta_1 \\
 \Theta_1&=& \mu_a \xi^a - \frac{1}{2} f_{ab}^c p_{c} \xi^a
    \xi^b
  \end{eqnarray*}
with $\xi^a$ being coordinates on $\g[1]$ and $p_{a}$ their $deg=-2$ conjugates.

The key point is that this $BFV_{-1}$ space can be also endowed with a BV
structure as follows. Let us choose a volume form $\lambda$ on $N$ and denote with $\nu_\lambda$ the corresponding berezinian; let 
$\nu_{gh}= c \ d\xi^1\ldots dp_1\ldots$ be a berezinian for the ghost part ($c\in\R$) with BV-laplacian 
$\Delta_{gh}=\frac{\partial^2}{\partial p_a \partial \xi^a}$. The total $BV$-laplacian is
$\Delta =\Delta_\lambda + \Delta_{gh}$. The QME for $S$ implies
$\Delta (S) = 0$, {\it i.e.} for each $a=1\ldots \dim \g$
\begin{equation} \label{eq: Delta mu}
 \Delta_\lambda(\mu_a) + f_{ab}^b = 0~~~.
\end{equation}
Let us introduce the fermionic delta function supported on $\mu^{-1}(0) \subset T^* [-1]N$ as
\begin{equation*}
\delta(\mu) =  \int\limits_{\fL_{gh}} e^{\frac{i}{\hbar}\Theta_1} \sqrt{\nu_{gh}} = c \ \mu_1 ... \mu_k
\end{equation*}
where $\fL_{gh} = \{
p_{a}=0 \} \subset T^*[-1]\g[1]$. Equation (\ref{eq: Delta mu}) is equivalent to the existence of a
volume form $\lambda_{red}$ on the quotient $N/G$ satisfying 
\begin{equation} \label{eq: lambda basic}
  i_{\delta(\mu)}\lambda = q^*\lambda_{red}   \;.
\end{equation}
The above equation defines $\lambda_{red}$: remark that it contains the dependence on the choice of the ghost berezinian $\nu_{gh}$ and on $\hbar$.
We conclude that, in this case,
($T^*[-1](N\times g[1]),\omega_{can},\nu_\lambda\times\nu_{gh},S$) is a BV-BFV manifold and the reduced $BV$ manifold is 
($T^*[-1](N/G),$ $\omega_{can},\nu_{\lambda_{red}}$).

Let  $F = \sum\limits_{r\leq0} F_r \in C(\fF)$ be concentrated in non positive $ga$ degree and be a quantum observable, 
{\it i.e.} $\Delta (F e^{iS/\hbar})=0$. This in particular implies that $f$ reduces to $f_{red}$ and
$\Delta_\lambda f + f^a_a=0$, where $F_0= f + f^a_a \xi^a p_a+\ldots $. %One immediately computes that
%$$
%\Delta_\lambda (e^{\frac{i}{\hbar}S_0} f \delta(\mu)) = 0~~~~.
%$$
The following gives a finite dimensional version of the computation of correlators for the $BV-BFV$ system.

\smallskip
\begin{proposition} \label{lma: integrals on F_red}
Let
$\tilde{\Gamma} \subset N/G$ be a submanifold such that there exists a $\Gamma \subset N$
  with $q|_\Gamma:\Gamma \to \tilde{\Gamma}$ a local diffeomorphism. Then, 
$$
 \langle F \rangle_\lambda = \int\limits_{N^*[-1]\Gamma \times \fL_{gh} } Fe^{\frac{i}{\hbar}S}
  \sqrt{\nu_\lambda \times \nu_{gh}} = \int\limits_{N^*[-1]\tilde{\Gamma}} 
  f_{red} e^{\frac{i}{\hbar}S_{red}} \sqrt{\nu_{\lambda_{red}}} = \langle
  f_{red} \rangle_{\lambda_{red}}
$$
yields the corresponding integral on $\fF_{red}$. Moreover, this
integral only depends on $f_{red}$ and the homology class of the gauge fixing
$\tilde{\Gamma}$.
\end{proposition}

{\it Proof}: The following identity yields
\begin{equation}\label{integration_ghost_gauge_fixing} 
\langle F \rangle=\int\limits_{N^*[-1]\Gamma \times \fL_{gh} }( F \
e^{\frac{i}{\hbar}S}) \sqrt{\nu_{\lambda} \times \nu_{gh}} = \int\limits_{N^*[-1]\Gamma}( f e^{\frac{i}{\hbar}S_0}\ \delta(\mu)) \sqrt{\nu_\lambda}~.
\end{equation} 
Recall that \cite{Sch} the BV laplacian $\Delta_\lambda$ on $T^*[-1]N$ can be obtained from
    de Rham differential $d$ on $T[1]N$ via a Fourier-like transform
    $\mathcal{F}_\lambda:C(T[1]N) \to C(T^*[-1]N)$ for which
\begin{equation*}
\int\limits_{N^*[-1]\Gamma} \mathcal{F}_\lambda(\alpha) \sqrt{\nu_\lambda} =
\int\limits_\Gamma \alpha ~,
\end{equation*}
 where $\alpha$ is a differential form.
Moreover, $\mathcal{F}^{-1}_\lambda(g)=i_g\lambda$ so that if $g$ is reducible, by using definition (\ref{eq: lambda basic}) we get 
$\mathcal{F}_\lambda^{-1}(g\delta(\mu))=q^*\mathcal{F}_{\lambda_{red}}^{-1}(g_{red})$.
The result then follows by applying it to 
$g=f\exp(iS_0/\hbar)\delta(\mu)$ in (\ref{integration_ghost_gauge_fixing}). $\square$

\begin{remark}
Notice that the 'gauge fixing' transversality condition on
$\Gamma$ with respect to the quotient $q:N \to N/G$ is
\emph{necessary} for $\langle F\rangle_\lambda$ to be different from zero due
to the presence of $\delta(\mu)$.
\end{remark}
\begin{remark} Recall the
    Weil model for equivariant cohomology $(T[1](N\times
    \g[1]),d_W)$ seen as a Q-manifold. Using a volume form $\lambda$ on $N$ satisfying
    $(\ref{eq: lambda basic})$ and an invariant inner product on $\g$, one can
    define a formal Fourier-like transform to the $BFV_{-1}$ space $\mathcal{F}:C(T[1](N\times
    \g[1])) \to C(T^*[-1](N\times
    \g[1]))$. It sends the Weil differential $d_W$ to a
  laplacian $\Delta^{BFV}$ given by
  $\Delta^{BFV}=\Delta_\lambda + \frac{\partial^2}{\partial p_a \partial \xi^a} +
  \{\Theta_1, \cdot \}$ where $\Theta_{1}$ is the pure BFV
  charge given above. This establishes a clear relation
  between the Weil model for equivariant de Rham cohomology together with integration of basic forms on the one hand and the
  $BFV_{-1}$ model as a BV-manifold together with integration of the observables
  given in the above Lemma, on the other.
\end{remark}

\begin{remark} %\textbf{Here we comment to \cite{Zuc10}. Still to do.}
Similar constructions to the BV-structure on $\fF$ presented in this
section, were considered in \cite{Zuc10} for the cases of hamiltonian
group actions on BV-algebras. On the other hand, in that reference the focus is on
the associated $\g$-invariant cohomology rather than on the
\emph{BFV-cohomology} which is the one that leads to functions on the
reduced space
$\fF_{red}$ as studied here.
%
%For some zero modes AKSZ theories and when there is a (hamiltonian)
%Lie group action involved, in \cite{Zuc10} it was studied how to introduce BV algebra structures (i.e. compatible BV laplacians) on the corresponding
%extended BFV systems. BUT WHEN HE STUDIES 'GAUGE INVARIANT' OBSERVABLES, HE IS NOT INTERESTED IN THE SAME
%COHOMOLOGY AS WE ARE (I.E. NOT THE REDUCED COHOMOLOGY)
\end{remark}
\bigskip

%%%%%%%%%%%%%%%%%%%%%%%%%%%%%%%%%%%%%%%%%%%%%%%%%%%%%%%%%%%%%%%%%%%%%%%%%%%%%%%%%%%%%%%%%%%%%%%%%%%%%%%%%%%%%%%%%%%%%%%%%%%%%%%%%%%%%%%%%%%%%%%%%%%%%%%%%%%%%%%%

\bigskip
\bigskip

%%%%%%%%%%%%%%%%%%%%%%%%%%%%%%%%%%%%%%%%%%%%%%%%%%%%%%%%%%%%%%

\subsection{General features}\label{general_BFV_AKSZ}

A $BFV_n$ space can be used as a target for the AKSZ construction. In
fact a $BFV_{n}$ space is, in particular, a degree $n$-symplectic manifold 
endowed with an homological hamiltonian of degree $n+1$, which is the
piece of data needed for the \emph{target space} of the AKSZ construction of an
$n$-dimensional topological field theory. We call it the $BFV-AKSZ$ model. By
construction, the
corresponding space of fields $\fF$ will inherit both BFV and BV
structures and becomes a $BV-BFV$ manifold as described in Section \ref{BFV-BV}.

In this section we discuss some general features of the BFV-AKSZ
model; the discussion will be rather formal at the field space level and some of the statements will become 
precise once that we discuss finite dimensional examples. The main point is to sketch how this BFV-AKSZ
system can be used as a model for the AKSZ system defined by the underlying
\emph{reduced target geometry} $\M_0//C_{\Theta_1}$. Namely,
reduction-data-compatible computations using the BFV-AKSZ system
can be used to obtain the corresponding computations in the reduced theory.

Let $({\cal M},\Omega_n,\Theta)$ be the $BFV_n$ space with $\Theta=\sum\limits_{r\leq 1}\Theta_r$ defining the regular degree $n$-reduction data
$({\cal M}_0,\Omega_n|_{{\cal M}_0},C_{\Theta_1},\theta=\Theta|_{{\cal M}_0})$ be  as in Lemma \ref{fromBFVtoReduction}.
We shall keep the notation involving the reduction data from Section \ref{BFV-AKSZ system}. Consider the BV field spaces defined by
$$
\mathfrak{F}_0 = {\rm Map}(T[1]\Sigma _{n+1},{\cal M}_0)  ~,~~~~~~ \mathfrak{F}_{red} = {\rm Map}(T[1]\Sigma _{n+1},{\cal M}_{0}~//C_{\Theta_1})
$$
$$
\mathfrak{F}={\rm Map }(T[1]\Sigma_{n+1},{\cal M})~.
$$
The field spaces above are endowed with anti-brackets of degree $+1$
and with homological charges given by the 
AKSZ actions $S_\theta$, $S_{\theta_{red}}$ and $S_{\Theta}$,
respectively, as described in section \ref{Section_AKSZ}. We shall proceed formally by treating them as finite dimensional manifolds
(this will indeed be the case for zero modes AKSZ theories \cite{BMZ} discussed in Subsection \ref{zero_modes_PSM} and
Appendix \ref{app: zero modes AKSZ commutes BFV}).

The choice of local Darboux coordinates $(x^\mu,p_{\mu},\xi^a,p_a)$ on ${\cal M}$, as described in Section \ref{BFV-AKSZ system}, defines the 
superfields $(\superx^\mu,\superp_{\mu},\superxi^a,\superp_a)$. In
particular we call the components of $\superxi$ and $\superp$ the (BFV) {\it ghosts} and {\it antighosts}. 
The coisotropic $C_{\Theta_1}\subset {\cal M}_0$ gives rise to what can be called a first class constraint AKSZ system on $\mathfrak{F}_0$, 
by considering the coisotropic submanifold
$\mathfrak{C}_{\Theta_1}={\rm Map}(T[1]\Sigma_{n+1},C_{\Theta_1})\subset \mathfrak{F}_0$. It can be defined as the vanishing of all the 
components $\Theta_{1,aI}$ of
\begin{equation*}
 \Theta_{1,a}(\superx,\superp)=ev^*\Theta_{1,a} = \sum_I \Theta_{1,aI} \theta^{I}
\end{equation*}
expanded as in eq. $(\ref{eq: ev*f})$. Moreover, if the symplectic potential 1-form $\vartheta_{{\cal M}_0}$ on ${\cal M}_0$ is also 
reducible by $C_{\Theta_1}$ one can say
that {\it AKSZ\ commutes with reduction} in the sense that
\begin{equation}\label{lemma: general AKSZ commutes with reduction}
\mathfrak{F}_0//\mathfrak{C}_{\Theta_1}=\mathfrak{F}_{red}
\end{equation}%
together with their QP-structure. In particular, $S_{\theta}|_{\mathfrak{C}}=\tilde{q}^*S_{\theta_{red}}$ where $\tilde{q}:\mathfrak{C}_{\Theta_1} \rightarrow \mathfrak{F}_0//\mathfrak{C}_{\Theta_1}$
denotes the quotient map.

%%%%%%%%%%%%%%%%%%%%%%%%%%%%%%%%%%%%

%
Moreover, one could formally repeat the $BFV_{-1}$ construction of
section \ref{BFV-BV} to the reduction data $(\mathfrak{F}_0,\mathfrak{C}_{\Theta_1},
{\boldsymbol \omega}_{-1},S_{\theta})$ at the field space level and
show that {\it AKSZ commutes with BFV}. By this we mean that the
construction in which one first builds the $BFV_n$ for the target and
then applies the AKSZ construction, namely $\fF$, is a $BFV_{-1}$
model for the above reduction data on $\fF_0$. This statement can be
made precise in the finite dimensional setting of zero modes AKSZ
theories, see Appendix \ref{app: zero modes AKSZ commutes BFV} for
details.

%\marginpar{{\bfseries A*}}
%%%%%%%%%%%%%%%%%%%%%%%%%%%%%%%%%%%%

\medskip

Let us now analyze the BV structure of the system defined by $S_{\Theta}=\sum\limits_{r\leq 1} S_{\Theta_r}$.

%%%%%%%%%%%%%%%%%%%%%%%%%%%%%%%%
%\begin{remark} {\bf this is probably to remove}
%The BV laplacian acting on ${\cal F}_{ext}$ is of the form $\Delta_{ext}=\Delta+\Delta_{ghost}$, where $\Delta_{ghost}$ acts on the ghost part.
%Let $\mathfrak{I}\in C(\fF)$ be the vanishing ideal of $\fC \subset \fF$,
%generated by the components $\mu_{a_l,I}^{(l)}(u)$. We want that the BV laplacian $\Delta$ acting on $\mathfrak{F}$ be
%\emph{reducible by} $\fC$, in the sense that $\Delta \fI \subset \fI$; 

%{\bf This condition is too strong. it implies that all $\mathfrak{F}$ is invariant}:

%then one can check that it induces a BV
%laplacian $\Delta_{red}$ on the reduction $C(\fF // \fC)=(C(\fF)/\fI)^\fI$ via $\tilde{q}^*(\Delta_{red} F_{red}) = (\Delta)|_{\fC}$ for $F$
%reducible and $\tilde{q}:\fC \rightarrow \fF//\fC$ denoting the quotient map. Motivated by this, we assume the (in general stronger) condition that
%\begin{equation}\label{reducibility of Delta}\Delta \mu_{a_l,I}^{(l)} = 0\end{equation}

%The condition that $S_{ext}$ solves the quantum master equation, {\it i.e.} $\Delta_{ext}S_{ext}=0$ expanded in the antighost fields gives
%$\Delta S_{ext}^r + \Delta_{ghost} S^{r+1}_{ext}=0$. The condition for $r=0$, $\Delta S + \Delta_{ghost}S^1_{ext}=0 $, together with (\ref{reducibility of Delta}),
%implies the quantum master equation for $S_{red}$ only if $\Delta_{ghost}S^1_{ext}\in\mathfrak{I}$.  We will discuss more concretely this point in the
%examples in the following section.

%\end{remark}

%%%%%%%%%%%%%%%%%%%%%%%%%%%%%%%%%%%%%%%%%%%%%%%%%%%%%%%%%%%%%%
\smallskip

In the case $\fF = \fF_0 \times \fF_{ghosts}$, for example, when $\M$ comes
from the BFV construction of sect. \ref{BFVconstruction}, let us
consider a BV laplacian $\Delta$ on $C(\fF)$ given by
$$
\Delta = \Delta_0 + \Delta_{gh}
$$
where $\Delta_0$ is a BV laplacian on $\fF_0$ and $\Delta_{gh} \simeq
\frac{\partial^2}{\partial \superp_a \partial \superxi^a}$ is a
laplacian acting on the ghost variables. For general $\fF$, we require
that $ga(\Delta)=0$ as in the above case. We know that $S_\Theta \in C(\fF)$
satisfies the CME by construction and we shall further assume that the QME holds
so that $\Delta(S_\Theta)=0$.  Notice that this last equation can be expanded in
$ga$-degrees in which each term has to be zero. In the toy model of
section \ref{BFV-BV}, we saw that the $ga=1$ term implied the
reducibility of the volume form from $\fF$ to $\fF_{red}$.

Let us now consider a gauge fixing $\mathfrak{L}\subset{\rm Map}(T[1]\Sigma_{n+1},{\cal M_>})$, {\it i.e.} in the ghost sector we
put to zero the antighosts $\superp_{a}=0 $. With this choice
$ S_\Theta|_{\mathfrak{L}} =S_\theta-\int_{T[1]\Sigma } \superxi^a\Theta_{1,a}(\superx,\superp)$, so that the ghosts $\superxi$ appear as 
Lagrange multipliers for the constraints $\Theta_{1,aI}$ on $\fF_0$. Notice that
when $\mathfrak{F}=\mathfrak{F}_0\times\mathfrak{F}_{ghosts}$, 
the ghosts and antighosts can be 
interpreted as ultraviolet degrees of freedom in the spirit of the
effective action described in Section \ref{Section_AKSZ}. The UV gauge fixing $\mathfrak{L}_{ghost}=\{ \superp_a
=0 \}$ thus yields
as the effective action 
$$
e^{\frac{i}{\hbar}S_{eff}} = \int\limits_{\mathfrak{L}_{ghost}} e^{\frac{i}{\hbar}S_\Theta} = e^{\frac{i}{\hbar}S_\theta} \delta(\Theta_{1,a}(\superx,\superp))\;,
$$
{\it i.e.} the AKSZ action restricted to the constraint
$\mathfrak{C}_{\Theta_1}$.

\begin{remark} \label{rmk: reduction by delta}
Recall that for a coisotropic ideal $I \subset C(\fF_0)$, the Poisson
reduction is defined as $C(\fF_{red}):= N(I)/I$ where $N(I)=\{ f\in
C(\fF_0): \{I,f \} \subset I \}$ are the reducible functions. Then,
for $f \in N(I)$, one formaly has that $f
\delta(\Theta_{1,a}(\superx,\superp)) \equiv f_{red}$ since
multiplying by the delta
function is equivalent to perform the quotient by $I$. In particular, for the effective
action above we have $S_{eff} \equiv S_{\theta_{red}}$.
\end{remark}

%%%%%%%%%%%%%%%%%%%%%%%%%%%%%%%%%%%%%%%%

%%%%%%%%%%%%%%%%%%%%%%%%%%%%%%%%%%%%%%%%%%%

%Observables
Let $F=\sum\limits_{r\leq 0} F_r\in C({\cal M})$ be a classical observable concentrated in nonpositive $ga$-degree, for instance one in the image 
of the map (\ref{eq: general ag=0 BFV cohomology}). We recall that $f=F|_{{\cal M}_0}$ is a reducible observable that reduces to 
$f_{red}\in C({\cal M}_0//C_{\Theta_1})$. Denoting $O_F=F(\superx^\mu,\superp_\mu,\superxi^a,\superp_a)$; we then have that
\begin{equation} \label{eq: F_ext on L_ghost}
O_F|_{\mathfrak{F} \times \mathfrak{L}_{ghost}}=O_f=f(\superx,\superp) \in C(\mathfrak{F}) \;.
\end{equation}

Assuming that $\Delta O_F=0$, let us compute the expectation values on the extended BV system,

\begin{eqnarray}
\nonumber \left\langle O_F \right\rangle _{\cal L} &=&\int\limits_{\mathfrak{L}} O_F \ e^{\frac{i}{\hbar}S_{\Theta}} \\
\nonumber &=&\int\limits_{\mathfrak{L}_0\times \mathfrak{L}_{ghost} } O_f \ e^{\frac{i}{\hbar}S_\theta-i\int\limits_{T[1]\Sigma }\superxi^{a}\Theta_{1,a}(\superx,\superp)} \\
\label{eq: extended exp val computation} &=& \int\limits_{\mathfrak{L}_0}  \ \delta (\Theta_{1,a}(\superx,\superp)) \ O_f \ e^{\frac{i}{\hbar}S_\theta} \\
 &=:&\nonumber \left\langle O_{f_{red}}\right\rangle_{red}
\end{eqnarray}%
where $\fL_0 \subset \fF_0$ is an IR gauge fixing.

In the toy model of Section \ref{BFV-BV}, we showed that the last line above
is indeed an equality. In this
 general infinite-dimensional setting, the last line above must be
 seen as a definition of the expectation value for the theory with reduced target
$(M_{n}//C_{n},\omega _{M_{n}//C_{n}},\theta
_{M_{n}//C_{n}})$. %As in Sections \ref{Section_AKSZ} and \ref{BFV-BV}, condition $\Delta O_F=0$ is needed for the
%independence of the above integral on the gauge fixing choice. 
Also notice that the above correlator actually depends on the 
class of $F$ in $H_Q$ which, for $\Theta_0\not = 0$, can correspond to
different classes in $H_{Q_{red}}$ since the map
$\lambda(f_{red})=F^\infty(f)$ of Proposition \ref{prop: map from reduced
  cohomology} can be non injective (see the discussion in 
Remark \ref{obstructiontocohomology}).
%%%%%%%%%%%%%%%%%%%%%%%%%%%%%%%%%%

%%%%%%%%%%%%%%%%%%%%%%%%%%%%%%

From the finite dimensional computation in section \ref{BFV-BV}, we
also notice that the freedom of independently choosing
the infrared and ultraviolet gauge fixing is apparent. Indeed the lagrangian $\mathfrak{L}_0$
must be transversal to the gauge transformations. This is a point that
will be analyzed further in the example of the following Section for
the PSM.

\smallskip

\begin{remark}
The case $n=0$ corresponding to the usual $BFV_{0}$ construction is discussed in Appendix \ref{app: ham syst}. It is
shown how, with a slight modification in order to encode non-topological dynamics, the corresponding BFV-AKSZ system yields a BV-quantization procedure
of hamiltonian systems with symmetry (see also \cite{GD99}).
\end{remark}

%%%%%%%%%%%%%%%%%%%%%%%%%%%%%%%%%%%%%%%%%%%%%%%%%%%%%%%%%%%%%%%%%%%%%%%%
%%%%%%%%%%%%%%%%%%%%%%%%%%%%%%%%%%%%%%%%%%%%%%%%%%%%%%%%%%%%%%%%%%%%%%%%%

\section{Reduction by group actions for $n=1$ }\label{groupactionn_1}

%%%%%%%%%%%%%%%%%%%%%%%%%%%%%%%%%%%%%%%%%%%%%%%%%%%%%%%%%%%%%%%%%%%%%%%%%%%%%%%%%%%%%%%%%%%%%%%%%%%%%%%%%%%%%%%%%%%%%%%%%%%%%%%%%%%%%%%%%%%%%%%%%%%

\subsection{$BFV_1$ for Poisson actions}
\label{subsec: BFV_1 for G acts} 
We discuss in this section examples of the $BFV_1$
construction outlined in the previous sections for the case of a group
action. Nonnegatively graded symplectic manifolds of 
degree $1$ encode Poisson geometry, so that the geometric background
of this Section is the reduction of Poisson structures by group
actions\footnote{Notice the difference with the negatively
  graded $T^*[-1]N$ of
 section \ref{BFV-BV} in which one focuses on degree
  \emph{zero} charges $S_0$ instead of degree \emph{two} charges
  (i.e. bivectors) as in the case of $T^*[1]N$.}. Relevant
facts of Poisson Lie geometry are summarized in Appendix \ref{subsec: Basics Poisson actions}.
We will consider first the reduction by the Poisson action of a Poisson Lie group, which is encoded by a degree $1$ momentum map. 
As a final example, we will consider also degree $0$ momentum map 
in the case of an invariant Poisson structure.

Let $(M,\pi_{M})\ $ be a Poisson manifold and let $(G,\pi_{G})\ $ be a Poisson Lie group acting on it with a Poisson action
(see Appendix \ref{subsec: Basics Poisson actions} for the basic notions on Poisson reduction). 
Analogous to section \ref{BFV-BV}, choosing a basis $\{T_a\}$ of $\g$ and graded coordinates $(x^{i},b_{i})$ on $T^{\ast }[1]M$, the 
fundamental vector fields $v_a=v_a^i\partial_i$ 
define the (degree $1$) equivariant moment map $\mu^{(1)}_G:T^*[1]M \to \mathfrak{g}^*[1]$ for the lifted $G-$action on $T^*[1]M$,
%$(\mu_G^{(1)})_a\in C^1(T^*[1]M_0)$, 
whose components are
\begin{equation}
 (\mu_G^{(1)})_a=v_{a}^{i}b_{i} \in C^1(T^*[1]M).
\label{eq: moment map for lifted T*[1] action}
\end{equation}% 

Since the action is Poisson, then $\pi_{M}\in C^2(T^*[1]M)$ is 
not invariant but satisfies
\begin{equation}
\{(\mu^{(1)}_G)_{a},\pi_{M}\}=\frac{1}{2}\tilde{f}_{a}^{bc}(\mu^{(1)}_G)_b(\mu^{(1)}_G)_c~,
\label{eq: PL action and pi}
\end{equation}%
where $\tilde{f}_{a}^{bc}$ are the dual structure constants for $\mathfrak{g}%
^{\ast }$. Let us consider the degree $1$ reduction data $(T^{\ast
}[1]M,\omega=dx^idb_i ,C_{1}=\mu^{(1)}_G{}^{-1}(0),\pi_{M_0})$. If
the $G$-action is free and proper then the reduced manifold is
\begin{equation*}
T^{\ast }[1]M//C_1=T^{\ast }[1](M/G)~~~~~.
\end{equation*}%
From eq. $(\ref{eq: PL action and pi})$, it follows that $\pi_{M}$ is reducible to a
bivector $\pi_{M/G}$ on $M/G$.

The $BFV_{1}$ construction applied to the above reduction data gives
\begin{equation}
\mathcal{M}:=T^{\ast }[1]M\times T^{\ast }[1]\mathfrak{g}[1], \label{eq: BFV_1 space for G-action}
\end{equation}%
with the symplectic form $\tilde{\omega}=\omega \oplus \varpi$
where $\varpi $ is the canonical degree $1$ symplectic structure on $T^{\ast
}[1]\mathfrak{g}[1]=\mathfrak{g}[1]\times \mathfrak{g}^{\ast }[0]=T^{\ast
}[1]\mathfrak{g}^{\ast }$. 
Let us denote with $(\xi^a,p_a)$ ghosts and antighosts respectively; the $BFV_{1}$ charge starts with
\begin{equation}
\Theta_{\pi_{M}} =\pi_{M}+(\mu^{(1)}_G)_{a}\xi ^{a}-\frac{1}{2}f_{bc}^{a}p_a\xi ^{b}\xi ^{c}+\ldots\in C^{2}(\mathcal{M})
\label{eq: BFV_1 charge G-action}
\end{equation}%
When $\pi_{M}$ is directly invariant (case $\tilde{f}^{bc}_a=0$), no
more terms are needed above. In the general case, though, due to \emph{Poisson-Lie} invariance of $\pi_{M}$, corrections are needed in order 
to satisfy the master equation. If the constraints are regular, 
from Proposition \ref{prop: map from reduced cohomology}, such a charge exists at least as a formal
power series on the antighosts $p_{a}$. We are going to show that it always exists. Let $e:\g^*\rightarrow G^*$ 
be a local diffeomorphism around $0$, such that $e(0)=e\in G$; then one has
the lift $\tilde{e}:T^*[1]\g^*\rightarrow T^*[1]G^*$; it is defined on the coordinates $x^{\alpha},\beta_\alpha$ of $T^*[1]G^*$ as
\begin{equation}\label{exponential}
\tilde{e}^*(x^\alpha) = e^\alpha(p) \;, ~~~~ \tilde{e}^*(\beta_\alpha)=\frac{\partial e_a^{-1}}{\partial x^\alpha} \xi^a
\;.\end{equation}
We assume that $x^\alpha$ is zero on the identity $e_{G^*}\in G^*$.
Let us define 
$$ \Theta_{\pi_{M}}= \tilde{e}^*(\pi_{M\vartriangleleft G^*}) ~,~~~~$$ 
where $\pi_{M\vartriangleleft G^*}=\pi_{M}+ \pi_{G^*} + (\mu_G^{(1)})_a k^a$ is the semidirect Poisson structure on 
$M\times G^*$, (see \ref{semi_direct_prod} for notations).

\begin{proposition}\label{BFV_poisson_lie}
$\Theta_{\pi_{M}}$ is a $BFV_1$ charge for the reduction data $(T^*[1]M,\omega_1,\mu_G^{(1)}=0,\pi_{M})$.
\end{proposition}
{\it Proof}. Since $\tilde{e}$ is a symplectomorphism, $\Theta_{\pi_{M}}$ is homological. Let us check that 
$\Theta_{\pi_{M}}$ is negatively $ga$-graded and has the expansion of eq. 
(\ref{eq: BFV_1 charge G-action}). Keeping in mind the dependence on $\xi$ of (\ref{exponential}) and that $\pi_{G^*}(0)=0$, we easily
get that $\Theta_{\pi_{M}}=\sum\limits_{r\leq 1} \Theta_r$, where $\Theta_r$ is the component of $ga$-degree $r$. Moreover, we compute
that
$\Theta_1 = (\mu_G^{(1)})_a\tilde{\xi}^a +\frac{1}{2} f_{ab}^c
\tilde{\xi}^a\tilde{\xi}^b \tilde{p}_c$ and
$\Theta_0 = \pi_{M}+\tilde{f}^{ac}_b (\mu_G^{(1)})_a\tilde{p}_c\tilde{\xi}^b + O(p^2)$, 
where $\tilde{\xi}^a=k^{a\alpha}e_{b\alpha}\xi^b$ and $\tilde{p}_c=k_{c\gamma}e^{\gamma d}p_d$ defines a symplectomorphism.
$\square$   

In general $e$ can be chosen as $exp$, the exponential map, which is in general just a local diffeomorphism around the identity.
The construction still works if we restrict ourselves to this neighbourhood of the
identity, since this is the region that is relevant
for BFV-cohomology. Recall also that when $G=K$ is compact and simple then $%
G^{\ast }$ is globally diffeomorphic to $\mathfrak{g}^{\ast }$
(the $AN$ part of the Iwasawa decomposition of $K^{\mathbb{C}}$), see \cite{GW}.

For AKSZ\ applications of the next sections, we will use this
non-linear version of the BFV model,
namely, we shall work with the space $exp(BFV^\theta_{1}):=T^{\ast }[1](M\times
G^{\ast })\ $ endowed with the $BFV$ charge $\pi _{M\vartriangleleft G^{\ast }}$.

%\smallskip
%\begin{remark} \label{rmk: reduced cohomology G-Poisson case} {\bf keep it if useful}. {\rm (Vertical complex)}.
%The vertical complex introduced in Remark \ref{vertical_complex} for the above geometry is
%$(\mathcal{V^{\cdot }},d_v)=(\Lambda \mathfrak{g}^*\otimes \Gamma_{M_0}(\Lambda(TM_0/\mathfrak{g}_{M_0})),d_{CE}+d_{\pi_{M_0}}^v+\xi^a l_a)$
%where $\mathfrak{g}_{M_0} \subset TM_0$ is the vector subbundle spanned by the inifinitesimal generators
%$v_a$, $d_{CE}$ is the Chevalley-Eilenberg differential on the ghosts $\xi^a \in \Lambda \mathfrak{g}^*$ and $d_{\pi_{M_0}}^v$, $l_a$
%are induced from $\{\pi_{M_0}, \ \}$ and $\{v_a, \ \}$, respectively, being $\{,\}$ the bracket on $C(T^*[1]M_0)$. %{\bf Question: CHECK IF,
%when $\mathfrak{g}$ is a \emph{reductive} Lie algebra, the map $q^*$ from reduced cohomology to the cohomology of the
%vertical complex is injective and thus, that the map $(\ref{eq: general ag=0 BFV cohomology})$ is injective.
%APPENDIX A OF \cite{Zuc10}.  THIS SHOULD WORK ALSO FOR ANY $M_n=T^*[n]N$ WHERE THE ACTION IS LIFTED FROM $N$
%AND LEAVES THE CHARGE IN $M_n$ INVARIANT...}
%\end{remark}

%%%%%%%%%%%%%%%%%%%%%%%%%%%%%%%%%%%%%%%%%%%%%%%%%%%%%%%%%%%%%%%%%%%%%%%%%%%%%%%%%%%%%%%%%%%%%%%%%%%%%%%%%%%%%%%%%%%%%%%%%%%%%%%%%%%%%%%%%%%%%%%%%%%%%%%%%%%%%%%%
\medskip 

\begin{remark}
One way of avoiding these ``non-linearities", namely the extra terms in the
modified charge or the non flat $G^{\ast }$ factor in the exponentiated
version, could be to go one degree higher to $n=2$. In this case, one can encode
the Poisson-Lie action information in a Dirac structure \cite{BCS} inside a
Courant algebroid as it will be done in subsection \ref{subsec: BFV_2 for poisson
actions}. 
\end{remark}

\medskip

As a final example, let us consider also a momentum map in degree 0
for the $G$-action. Let $G$ act on $(M,\pi_{M})$ by Poisson
diffeomorphisms, 
{\it i.e.} the Poisson tensor $\pi_{M}$ is $G$-invariant (case
$\tilde{f}^{ab}_c=0$) and assume that this action is hamiltonian. This
means that there is a (degree $0$) equivariant moment
map $\mu^{(0)}_G:M\longrightarrow \mathfrak{g}^{\ast }$ such that
the fundamental vector fields are hamiltonian: $v_a=\pi_{M}
(d (\mu_G^{(0)})_a)$.
We shall consider the constraints defined
by the momentum maps $\mu^{(0)}_G$ and $\mu^{(1)}_G$ altogether, by taking
$C_1=\{\mu^{(0)}_G=0, \mu^{(1)}_G=0 \} \subset T^*[1]M$. In the
regular case, the
corresponding reduction yields
$$
T^*[1]M//C_1 =T^*[1](\mu^{(0)^{-1}}_G(0)/G)~,
$$ 
whose functions are the multivectors on the degree $0$ reduced space $\mu^{(0)^{-1}}_G(0)/G$. The $BFV_{1}$ construction gives
$$
{\cal M}=T^{\ast }[1](M\times \mathfrak{g}[1]\times \mathfrak{g}^{\ast }[-1])
$$
%\begin{eqnarray*}
%BFV_{1}(T^{\ast }[1]M_{0},\omega ,\left( \mathfrak{g}_{M_0}|_{C_{0}}\right)
%^{0}[1]) &=&T^{\ast }[1](BFV_{0}(M_{0},\pi,C_{0})) \\
%&=&T^{\ast }[1](M_{0}\times \mathfrak{g}[1]\times \mathfrak{g}^{\ast }[-1])
%\end{eqnarray*}%
with $BFV_{1}$ charge given by 
\begin{equation}
\label{eq: BFV_1 charge for moment map g*}
\Theta=\Theta_0 + \Theta_1 ~~~,
\end{equation}
%the cotangent lift\footnote{%
%Using a formula similar to the one used in $\left( \ref{eq: moment map for
%lifted T*[1] action}\right) $.} of the $Q$-vector field $%
%Q_{0}=\{\Theta _{0},-\}_{\pi \oplus \varpi^{-1} _{0}}$ on $BFV_{0}$ (recall
%eq. $\left( \ref{eq: BFV_0 charge}\right) $) to the function on $T^{\ast
%}[1]BFV_{0}$ 
where
$$
\Theta_0= \pi_M + p_a^{(1)} \xi^a_{(0)}~,~~~~~\Theta _{1} =(\mu^{(0)}_G)_{a}\xi^a_{(0)}+ (\mu^{(1)}_G)_a\xi^{a}_{(1)}+%
\frac{1}{2}p_{c}^{(1)}f_{ab}^{c}\xi^{a}_{(1)}\xi^{b}_{(1)}
+f_{bc}^{a}p_{a}^{(0)}\xi^b_{(0)}\xi^{c}_{(1)} ~~~,
$$
and $x^{i},b_{i}$ are coordinates on $T^{\ast }[1]M$, $\xi^{a}_{(1)}\in
C^{1}(\mathfrak{g}[1])$, $\xi^a_{(0)}\in C^{2}(\mathfrak{g}[2])$ are the ghosts and $p_a^{(0)}\in C^{-1}(\mathfrak{g}^{\ast }[-1])$,
$p_{a}^{(1)}\in C^{0}(\mathfrak{g}^{\ast }[0])$ are the corresponding $T^{\ast }[1]$ conjugates.

\begin{remark}
\label{rmk: Zucchini models}
The BFV-AKSZ model corresponding to (\ref{eq: BFV_1 charge for moment map g*}) is the sigma model  
proposed by Zucchini in \cite{Zuc07,Zuc08} as Poisson-Weil sigma models and by Signori in \cite{Signori} under the name of JPSM.
In \cite{Zuc08} it is shown that a sector of the underlying
BV-cohomology is related to the \emph{equivariant} Poisson cohomology of $%
(M,\pi_{M})$. In Appendix \ref{App: M times G* and Poisson equivar
  cohomology} we clarify this relation, by relating the cohomology of
the target QP-manifold to Poisson equivariant cohomology.
\end{remark}
%\end{remark}

% %% REF TO SIGNORI
%\begin{remark} More generally, let $(M,\pi)$ be Poisson and $C_0
%  \subset M$ be a coisotropic submanifold given as $C_0=J^{-1}(0)$
%  for a map $J:M \rightarrow V$ into a vector space $V$. We can consider
%  $\mu^{(0)}=J$, $\mu^{(1)}_a=v_a=\pi(dJ_a)$ and $C_1=\{ \mu^{(0)}=0,
%  \mu^{(1)}_a=0 \} \subset T^*[1]M$ so that
%  $T^*[1]M // C_1=T^*[1](M//C_0)$. The corresponding AKSZ-BFV
%  model is a 2d sigma model called JPSM in \cite{Signori}. In another
%  direction, one can show that the manifold $N=M \times V$
%  inherits a Poisson bivector $\pi_N$ and that $E_0=Graph(J) \subset
%  N$ is a Poisson submanifold which is isomorphic to $M$. Defining $E_1=\{x \in M, \alpha \in
%  V, J(x)-\alpha=0 \} \subset T^*[1]N$ one gets $T^*[1]N // E_1 =
%  T^*[1]E_0$. The corresponding AKSZ-BFV model is a 2d sigma model
%  called CPSM in \cite{Signori}. Notice that this last model does not
%  contain explicitly the full reduction to $M//C_0$ and that in
%  \cite{Signori} the idea was to achieve this by imposing appropriate
%  boundary conditions on the fields.
%\end{remark}
%\marginpar{{\bfseries A*}}

\bigskip
\bigskip

%%%%%%%%%%%%%%%%%%%%%%%%%%%%%%%%%%%%%%%%%%%%%%%%%%%%%%%%%%%%%%%%%%%%%%%%%%%%%%%%%%%%%%%%%%%%%%%%%%%%%%%%%%%%%%%%%%%%%%%%%%%%%%%%%%%%%%%%%%%%%%%%%55

\subsection{General reduction for PSM}

\label{Section_Poisson_red}
We described in Section \ref{subsec: BFV_1 for G acts}
the $BFV_1$ construction associated to the Poisson quotient $(M,\pi_{M})\rightarrow (M/G,\pi_{M/G})$, where $(G,\pi _{G})$
is a Poisson Lie group acting on $(M,\pi_{M})$. We study here the corresponding BFV-AKSZ models.

Let $(G^{\ast },\pi_{G^*})$ be the dual Poisson-Lie group and recall the semidirect Poisson structure $%
\pi _{M\vartriangleleft G^{\ast }}$ on $M\times G^{\ast }$. According to what we discussed in Section
\ref{subsec: BFV_1 for G acts}, the graded manifold $T^{\ast }[1](M\times G^{\ast
})$ and hamiltonian $\Theta =\pi _{M\vartriangleleft G^{\ast }}$
is a nonlinear version of the $BFV_1$ space for these reduction data.
The BFV-AKSZ model associated to these data is the PSM of this semidirect structure.

In this section, we shall consider more elaborate gauge fixings
than in the general discussion of Sect. \ref{general_BFV_AKSZ} which
make use of the underlying Poisson-Lie geometry. We shall then argue that this PSM with target $(M\times G^{\ast },\pi
_{M\vartriangleleft G^{\ast }})$ can be used to compute correlators of PSMs with target
$M/H$ for any coisotropic subgroup $H$ of $G$. See the Appendix \ref{subsec: Basics Poisson actions} for notations and basic results concerning
Poisson reduction.

The formal argument goes as follows. First notice that the space of superfields is
the direct product of $BV$ manifolds $\fF=\mathrm{Map}(T[1]\Sigma ,T^{\ast
}[1]M)\times \mathrm{Map}(T[1]\Sigma ,T^{\ast }[1]G^{\ast })$. We can then
consider the superfields $\Lambda$ with values in $G^{\ast }$ as ultraviolet degrees
of freedom and those $\Phi$ with value in $M$ as infrared. We choose as ultraviolet
gauge fixing $\mathcal{L}_{gh}=\mathrm{Map}(T[1]\Sigma ,N^{\ast }[1]H^{\perp })$,
where $H^{\perp }$ is the connected subgroup (that we assume closed)
integrating the annihilator subalgebra $\h^{\perp }\subset \g^{\ast
}$ coming from the coisotropic $H\subset G$. Notice that when $H=G$ then $G^\perp=\{e\}$ and $N^{\ast
}[1]\{e\}=T^*_e[1]G^*=g[1]$. Thus, in this particular case, the above gauge fixing corresponds to put
$p_a=0$ in the correspondence (\ref{exponential}), and so to the ghost
gauge fixing discussed in Section \ref{general_BFV_AKSZ}.

The effective action thus yields
\begin{equation*}
e^{\frac{i}{\hbar}S_{\mathrm{eff}}}=\int\limits_{\mathcal{L}_{gh}}D\Lambda \ e^{\frac{i}{\hbar}S_{\pi _{M\vartriangleleft G^{\ast
}}}}=e^{\frac{i}{\hbar}S_{\pi _{M}}}\delta _{H}~,
\end{equation*}%
where
\begin{equation}
\delta _{H}=\int\limits_{\mathcal{L}}D\Lambda \ e^{\frac{i}{\hbar}S_{\pi _{G^{\ast
}}}+\frac{i}{\hbar}\int_{T[1]\Sigma }(\mu _{G}^{(1)})_{a}(\Phi)k^{a}(\Lambda)}=\int\limits_{\mathcal{L}%
}D\Lambda e^{\frac{i}{\hbar}\int_{T[1]\Sigma }(\mu _{H}^{(1)})_{A}(\Phi)k^{A}(\Lambda)}
\end{equation}%
is the \textit{delta function} enforcing the constraint $\mu^{(1)}_{H}\circ \Phi
=0$. In the above derivation we used the fact that $\pi _{G^{\ast
}}|_{N^{\ast }[1]H^{\perp }}=0$ due to coisotropy of $H^{\perp }$, $\{T_{A}\}
$ is a basis for $\h$ and $(\mu _{G}^{(1)})_{a}k^{a}|_{N^{\ast }[1]H^{\perp
}}=(\mu _{H}^{(1)})_{A}k^{A}$ since $H^{\perp }$ is a subgroup. The effective
action is then $S_{\pi _{M}}$ restricted to $\mu^{(1)}_{H}\circ \Phi
=0$ which, due to the Poisson-Lie invariance and using remark \ref{rmk: reduction by delta}, it is
equivalent to the reduced action 
$S_{red}=S_{\pi _{M/H}}$ on $\fF_{red}$.

This argument can be made precise in a finite dimensional setting. We will
do it by analysing the reduction on the $%
BV$-theory of zero modes in the next subsection. This analysis must be seen both as a more precise
statement of the above argument and as check of the full conjecture.

%\begin{remark} \label{rmk: quotient target space}
%On the target space $T^*[1](M \times G^*)$, one can compute similarly
%\begin{equation*}
%e^{\pi_{eff}}=\int_{N^{\ast }[1]H^{\perp }}\sqrt{\nu _{G^{\ast }}}e^{\pi _{M\vartriangleleft G^{\ast
%}}}=e^{\pi _{M}}\delta _{H}\in C(T^{\ast }[1]M)
%\end{equation*}%
%where
%\begin{equation*}
%\delta _{H}=((\mu^{(1)}_{H})_{1}\ldots (\mu^{(1)}_{H})_{\dim H})~\int_{H^{\perp }}D\lambda ^{a}\rho _{G^{\ast }}\det
%(k^{An})\;,
%\end{equation*}% 
%denotes the \textit{fermionic delta function} on $(\mu^{(1)}_{H})^{-1}(0)$.
%It is clear
%that $f_{1}\delta _{H}=f_{2}\delta _{H}$ if and only if
%$(f_{1}-f_{2})|_{\mu^{(1)}_H=0}=0$. 
%As in remark \ref{rmk: reduction by delta}, we thus get that the effect of integrating
%on the ultraviolet degrees of freedom is to perform reduction of
%reducible
%observables. In particular, the effective action on target space is thus $\pi_{eff}
%\equiv \pi_{M/H}$ as expected.
%\end{remark}

%%%%%%%%%%%%%%%%%%%%%%%%%%%%%%%%%%%%%%%%%%%%%%%%%%%%%%%%%%%%
%%%%%%%%%%%%%%%%%%%%%%%%%%%%%%%%%%%%%%%%%%%%%%%%%%%%%%%%%%%%
\subsection{Reduction for zero modes of PSM}\label{zero_modes_PSM}

We analyze here the zero modes of the PSM with target the Poisson
manifold $(M\times G^{\ast },\pi _{M\vartriangleleft G^{\ast }})$ and source $\Sigma _{g}$, the compact surface of genus $g$.
We know from \cite{BMZ} that
the zero modes of the PSM with target $(M,\pi _{M})$ are described by an
AKSZ construction, whose BV-space of superfields is $\fF_0^Z:=\mathrm{Maps}(X_{\Sigma
_{g}},T^{\ast }[1]M)$, where $X_{\Sigma _{g}}$ is the cohomology ring of $%
\Sigma _{g}$ seen as as sheaf over a point.
%Thanks to the nature of the AKSZ construction
%all the considerations done on the target space in the previous subsection transfer automatically
%in this setting.

We introduce a symplectic basis
$\{e^{I},e_{I}\}_{I}^{g}$ of $H^{1}(\Sigma _{g})$ with the ring structure
given by $e_{I}\wedge e^{J}=\delta _{I}^{J}s_{2}$, where $s_{2}$ is the
volume form normalized to $\int_{\Sigma _{g}}s_{2}=1$.

Let $\Phi =(\x,\e)$ be the superfields of $\fF_0^Z$. We then define the genus $g$ momentum map
\begin{equation}
\zeta_{G,a}^{(g)}=ev^*(\mu_G^{(1)})_{a}=v_{a}(\x)^{i}\e_{i}=\zeta_{a}^{(0)}+\zeta_{a}^{(I)}e_{I}+\zeta_{(I),a}e^{I}+\zeta_{a}^{(2)}s_{2}
\label{graded_momentum_map}
\end{equation}%
The hamiltonian vector fields of the components of $\zeta_{G}^{(g)}$ define
an action of a graded Lie algebra $\g^{(g)}$. The even part is $\g%
_{even}^{(g)}=\langle v_{a}^{(2)}\rangle \oplus \langle v_{a}^{(0)}\rangle =%
\g\ltimes \R^{\dim \g}[2]$, where $\g$ acts on $\R^{\dim \g}$ with the
adjoint action; the odd part is $\g_{odd}^{(g)}=\langle
v_{a}^{(I)},v_{a(I)}\rangle =\R^{2g}[1]$, where $\g$ acts on each copy of $\R%
\subset \g_{odd}^{(g)}$ with the adjoint action. The definition of $\zeta_{H}^{(g)}$ and $\h^{(g)}$ for any subgroup $H\subset G$ is obvious.

Formula (\ref{lemma: general AKSZ commutes with reduction}) within this setting means that the space of zero modes of the PSM with
target $(M/H,\pi _{M/H})$ is obtained as Marsden-Weinstein reduction of $\fF_0^Z$. One can prove it directly by introducing
coordinates of $M$ adapted to the $H$-action:

\smallskip

\begin{lemma}
\label{reduction_genusg}
\begin{equation*}
\fF_{red}^Z:=\mathrm{Map}(X_{\Sigma_g}, T^*[1](M/H)) = (\zeta_H)^{-1}(0)/\h^{(g)} \;.
\end{equation*}
\end{lemma}

\smallskip

Let us now consider the zero modes of PSM with target the semidirect product
Poisson structure on $M\times G^{\ast }$. Let $\Psi =(\Phi ,\Lambda )$ be
the superfields of $\fF^Z:=\mathrm{Maps}(X_{\Sigma
_{g}},T^{\ast }[1](M\times G^*))$. The BV-action of zero modes is
\begin{equation*}
S_{\pi _{M\vartriangleleft G^{\ast }}}=\int ds_{2}\ ev^{\ast }\pi _{M\times G^{\ast
}}=\int ds_{2}\left( \pi _{M}(\Phi)+\pi _{G^{\ast
}}(\Lambda)+v_{a}(\Phi)k^{a}(\Lambda)\right) ~~.
\end{equation*}%
We want to consider $\Lambda $ as UV-degrees of freedom and take $\mathcal{L}_{UV}=%
\mathrm{Map}(X_{\Sigma _{g}},N^{\ast }[1]H^{\perp })$ as UV-gauge
fixing. In what follows, we shall describe the relevant BV-structure on the
space of zero modes.
%We repeat again the same computation as in previous sections. 

Let us introduce a volume form $V_{M}=\rho_{M} Dx$ on $M$
and a volume form $V_{G^*}=\rho_{G^*} D\lambda$ on $G^*$; let $%
\nu_{M}=V_{M}\otimes V_{M}$ and $\nu_{G^*}=V_{G^*}\otimes V_{G^*}$ be the
corresponding berezinian integration on $T^*[1]M$ and $T^*[1]G^*$,
respectively, as in Section \ref{Section_AKSZ}. We denote with $\Delta_{\nu_{M}}$, $%
\Delta_{\nu_{G^*}}$ and $\Delta_{\nu_{M\times
G^*}}=\Delta_{\nu_{M}}+\Delta_{\nu_{G^*}}$ the corresponding $BV$-laplacians.
The choice of the
volume forms on $M$ and $G^{\ast }$ define the berezinians $\nu _{M\times G^*}^{(g)}$
on the corresponding spaces of zero modes $\fF^Z$ as well (see \cite{BMZ} for details).
Let us denote with $\Delta _{\nu _{M\times G^*}}^{(g)}$ the corresponding $BV$-laplacian and recall the following
formula valid for any $F\in C^\infty(T^*[1](M\times G^*))$
\begin{equation}\label{zero_modes_laplacian}
\Delta _{\nu _{M\times G^*}}^{(g)}O_F=2s_{2}(1-g) O_{\Delta _{\nu
_{M\times G^*}}F}\;,
\end{equation}
where $O_F=\int ds\ F(\Psi)$.
As a consequence, the obstructions for $S_{\pi_{M\vartriangleleft G^*}}$ to solve the quantum master equation
are the same as on the target space, namely, that $\pi _{M\vartriangleleft G^{\ast }}$ satisfies
\begin{equation*}
\Delta _{\nu _{M\times G^{\ast }}}e^{\pi _{M\vartriangleleft G^{\ast }}}=0\;.
\end{equation*}%
In turn, the above equation on target space means that $\pi
_{M\vartriangleleft G^{\ast }}$ is a unimodular Poisson
structure. This is equivalent to
\begin{equation}
\Delta _{\nu _{G^{\ast }}}\pi _{G^{\ast }}+(\Delta _{\nu
_{M}}(\mu^{(1)}_G)_{a})k^{a}=0~,~~~~~\Delta _{\nu _{M}}\pi _{M}-(\Delta _{\nu _{G^{\ast
}}}k^{a})(\mu_G^{(1)})_{a}=0\;\;\;\;\;.  \label{unimod_semi_direct_prod}
\end{equation}

By using the Poisson Lie property (\ref{PLaction}) for $\pi_{G^*}$ and the fact that
$\pi_{G^*}(e)=0$ we get from
the first equation in $(\ref{unimod_semi_direct_prod})$ that
$\Delta_{\nu_{M}}(\mu^{(1)}_G)_{a}+f_{ab}^b=0$. Thus, in particular,
there is an induced volume form $V_{M/G}$ on the
quotient $M/G$ 
as in Section \ref{BFV-BV}. As a consequence, in the case $H=G$, there is an
induced volume $\nu^{(g)}_{M/G}$ on the reduced zero modes field space
$\fF_{red}^Z =\mathrm{Map}(X_{\Sigma_g}, T^*[1](M/G))$. 
Notice that, for more general $H \subset G$, one has to impose additional
compatibility conditions to have an induced volume on reduced space
$M/H$, namely to ensure that
$\Delta_{\nu_{M}}(\mu^{(1)}_G)_{A}+f_{AB}^B=0$. Observe that in the case $G$ is compact semisimple and $H \subset G$ closed,
one always obtain the desired induced volume on $M/H$.

\begin{remark}
Equation $(\ref{unimod_semi_direct_prod})$ implies that $\Delta _{\nu_{M}}(\mu^{(1)}_G)_{a}={\rm div}_{V_{M}}v_a = -c_a\in\R$, 
$\Delta _{\nu _{G^{\ast }}}\pi _{G^{\ast }}=c_a k^a$, $\Delta _{\nu _{G^{\ast
}}}k^{a}={\rm div}_{V_{G^*}}k^a= - c^a\in\R$ and $\Delta _{\nu
_{M}}\pi _{M}= c^a (\mu_G^{(1)})_a$. Recall that 
if a vector field has constant non zero divergence then the volume form is exact; for instance if $M$ is compact
then $c_a=0$, {\it i.e.} $v_a$ is divergenceless, and $\pi_{G^*}$ is unimodular. 
\end{remark}

%%%%%%%%%%%%%%%%%

The effective action gives
\begin{equation}
e^{\frac{i}{\hbar}S_{eff}}=\int\limits_{\mathcal{L}_{UV}}\sqrt{\nu _{G^{\ast }}^{(g)}}e^{\frac{i}{\hbar}S_{\pi
_{M\vartriangleleft G^{\ast }}}}=e^{\frac{i}{\hbar}S_{\pi _{M}}}\delta (\zeta_{H}^{(g)})\;.
\end{equation}%
From the discussion in remarks \ref{rmk: reduction by delta} %and \ref{rmk: quotient target space}, 
the above effective
action can be considered to be equivalent to the reduced zero-modes action
$S_{\pi_{M/H}}$. 
Also above, notice that for $g\not=0$ the momentum map
$\zeta_{H}^{(g)}$ also contains even components so that the delta function must be considered as an
ordinary distribution\footnote{Notice that to get these even delta functions the $i$ factor on the integrand $e^{\frac{i}{\hbar}S}$
  becomes important.}.

Let us now focus on the case $H=G$ and discuss the gauge fixing of the infrared
degrees of freedom.

Let $f\in C^\infty(T^*[1]M)$ be reducible, {\it i.e.} it satisfies
\begin{equation}\label{reducible_observable}
\{(\mu_G^{(1)})_a,f\}= f_a^b (\mu_G^{(1)})_b ~~,~~~~ \{\pi_{M},f\} = f^b (\mu_G^{(1)})_b\;,
\end{equation}
so that it induces an $f_{red}\in C^\infty(T^*[1]M/G)$ on the
quotient $q:M\rightarrow M/G$.  Suppose we have $F\in C^\infty(T^*[1](M\times G^*))$ a corresponding extended observable satisfying
$F|_{T^*[1]M}=f$ and $\Delta_{\nu_{M\times G^*}}F=0$.
 Let
$\tilde{\Gamma}\subset M/G$, $\Gamma \subset M$ as in Lemma \ref{lma: integrals on F_red}. 

%\begin{remark}
%On the target space $T^*[1](M \times G^*)$, following
%section \ref{BFV-BV} one has
%\begin{eqnarray} 
%\langle F \rangle = \int_{N^{\ast }[1](%
%\Gamma\times \{e\})}\sqrt{\nu _{M\times G^{\ast }}}F e^{\pi _{M\vartriangleleft
%G^{\ast }}} & = &\int_{N^{\ast }[1]\Gamma}\sqrt{\nu _{M}}\ fe^{\pi _{M}}\delta
%_{G} \label{reduced_integral} \\ 
% &=& \int_{N^{\ast }[1]\tilde{\Gamma}}\sqrt{\nu _{M/G}}\
% f_{red}e^{\pi _{M/G}}  = \langle f_{red}\rangle_{red} \nonumber
%\end{eqnarray}%
%so the integral only depends on $f_{red}$ and the homology class of
%$\tilde{\Gamma}$. Recall that the 'gauge fixing' transversality condition of
%$\Gamma$ is a necessary condition for the integral to be different
%from zero, since otherwise the fermionic delta function will make the integral vanish.
%\end{remark}

On the full zero modes field space now,
as a consequence of
(\ref{zero_modes_laplacian}), we get that $O_F$ is a quantum
observable and let us consider as IR-gauge fixing the lagrangian submanifold $\mathcal{L}_{IR}=%
\mathrm{Map}(X_{\Sigma _{g}},N^{\ast }[1]\Gamma)$. As in section \ref{BFV-BV}, we obtain the following characterization of correlators.

\smallskip
\begin{proposition} With the notations above, the correlator yields
\begin{eqnarray*}
\langle O_{F}\rangle_{\nu_M} = \int\limits_{\mathcal{L}%
_{UV}\times \mathcal{L}_{IR}}\sqrt{\nu _{M}^{(g)}\nu _{G^{\ast }}^{(g)}}%
~O_Fe^{\frac{i}{\hbar}S_{\pi _{M\vartriangleleft G^{\ast }}}}&=&\int\limits_{\mathcal{L}_{IR}}\sqrt{%
\nu _{M}^{(g)}} O_fe^{\frac{i}{\hbar}S_{\pi _{M}}}\delta (\zeta_{G}^{(g)}) \\
&=& \int\limits_{ \{ X_\Sigma \to N^*[1]\tilde{\Gamma} \} }\sqrt{%
\nu _{M/G}^{(g)}} O_{f_{red}}e^{\frac{i}{\hbar}S_{\pi _{M/G}}} \\
&=&
\langle O_{f_{red}}\rangle_{\nu_{M/G}} ~~~.
\end{eqnarray*}%
It thus depends only on $f_{red}$ and on the homology class of $\tilde{\Gamma}$.
\end{proposition}
The proof can be done by considering a cover of adapted coordinates on the
principal bundle $M \to M/G$.

Notice that the transversality condition on $\Gamma$ with respect to
the quotient $q:M \to M/G$ is also a necessary condition for the above integral be different from zero, since
$\zeta_G^{(g)}|_{{\mathcal L}_{IR}}= ev^*\mu^{(1)}_G|_{N^*[1]\Gamma}$.

\bigskip\bigskip

%%%%%%%%%%%%%%%%%%%%%%%%%%%%%%%%%%%%%%%%%%%%%%%%%%%%%%%%%%%%%%%%%%%%%%%%%%%%%%%%%%%%%%%%%%%%%%%%%%%%%%%%%%%%%%%%%%%%%%%%%%%%%%%%%%%%%%%%%%%%%%%%%%%%%%%%%%%%%%%%

\section{Reduction of a group action for $n=2$}\label{AKSZ-BV-n_2}

We give in this Section the details for the $BFV_2$ construction in the case of a group action. Symplectic NQ manifolds of
degree $2$ encode the structure of Courant algebroid, so that the geometric background is about the reduction of Courant algebroids.
The main reference for this topic is \cite{BCG}; we sketch some facts
in Appendix \ref{red_courant_algbd}. We shall consider in particular 
the case of exact Courant algebroids, {\it i.e.} the reduction by a group action of the $n=2$ symplectic manifold $M_2=T^*[2]T^*[1]M_0$. 

Let $G$ act freely on $M$
with $v_a=v_{a}^{i}\partial _{i}\in \mathfrak{X}(M)$ being the
infinitesimal generators.
The $G-$action on $M$ lifts naturally to $%
T^{\ast }[1]M$ and this one, in turn, lifts naturally to the symplectic $\M_0=T^{\ast }[2]T^{\ast }[1]M$. This action is hamiltonian
with moment map%
\begin{eqnarray*}
\mu^{(2)}_G &:&T^{\ast }[2]T^{\ast }[1]M\longrightarrow \mathfrak{g}%
^{\ast }[2] \\
(\mu_{G}^{(2)})_a &=&v_{a}^{i}p_{x^{i}}-\partial
_{j}v_{a}^{k}\ b_{k}p_{b_{j}}
\end{eqnarray*}%
where $(x^{i},b_{j})$ are graded coordinates on $T^{\ast }[1]M$ and $%
(p_{x^{i}},p_{b_{j}})$ denote the corresponding conjugated variables on $%
T^{\ast }[2]T^{\ast }[1]M$, of degree $2$ and $1$
respectively, with respect to the canonical degree $2$ symplectic form $\omega
_{2}$. We will also consider the degree one moment map
$$
\mu^{(1)}_G: T^*[2] T^*[1]M\rightarrow \g^*[1] ~~~,~~~~~
(\mu^{(1)}_G)_a= v_a^i b_i~;
$$
the collection of moment maps $\mu_G=(\mu^{(2)}_G,\mu^{(1)}_G)$ satisfies the algebra  (\ref{constraints_courant_algebroid}), (for $\lambda=0$).
We consider the coisotropic submanifold $\C=\mu^{-1}_G(0)\hookrightarrow T^*[2]T^*[1]M$; the
symplectic reduction of these constraints gives
$$\M_0//\C=T^*[2]T^*[1](M/G)~~.$$
Following the general
procedure of Section \re{BFVconstruction}, we get for the $BFV_{2}$ space
$${\cal M} = T^*[2](T^*[1]M\times\g[1]\times\g[2])\;.$$
Let us denote with $\xi_{(2)}$ ({\it resp.} $\xi_{(1)}$) the ghosts associated to $\mu^{(2)}_G$ ({\it resp.} $\mu^{(1)}_G$), and
$p_{\xi^{(2)}}$ ({\it resp.} $p_{\xi^{(1)}}$) the momenta, we see by a direct computation that the $BFV_{2}$ charge is
\begin{equation}
\label{BFV2charge}
\Theta_1= (\mu^{(2)}_G)_a \xi_{(2)}^a +  (\mu^{(1)}_G)_{a} \xi_{(1)}^a -\frac{1}{2} f_{ab}^c\xi_{(2)}^a\xi_{(2)}^b p_{\xi_{(2)}^c} -
f_{ab}^c\xi_{(2)}^a\xi_{(1)}^b p_{\xi_{(1)}^c}\;.
\end{equation}
Observe that ${\cal M}$ is nonnegatively graded so that the $BFV_2$ hamiltonian $(\ref{BFV2charge})$ defines a Courant algebroid structure on 
$(T+T^*)(M\times\g^*)$. 

\subsection{Reduction of an exact Courant algebroid}

We want now to encode the reduction of exact Courant algebroids, as described in Appendix \ref{red_courant_algbd}. Let us consider the exact
Courant algebroid defined by the hamiltonian $\theta_H\in C^3(T^*[2]T^*[1]M)$, where
$\theta_H = p_{b_i}p_{x^i} - \frac{1}{6} H_{ijk}p_{b_i}p_{b_j}p_{b_k}$, with
$H$ being the closed three-form representing the Severa class. We suppose that the $G$-action is isotropic and trivially extended so
that there exist one forms $\lambda_a = \lambda_{a,i}p_{b_i}$ such that $\Phi = H + \sum_at_a\lambda_a$ is equivariantly closed. 
We slightly modify the above construction just by redifining the degree one constraints to
\begin{eqnarray*}
\mu^{(1)}_{G\lambda}&:& T^*[2] T^*[1]M\rightarrow \g^*[1] \\
(\mu^{(1)}_{G\lambda})_a&=& v_a^i b_i+\lambda_{a,i}p_{b_i}.
\end{eqnarray*}
The constraints $\mu_G=(\mu^{(2)}_G,\mu_{G\lambda}^{(1)})$ still close the algebra (\ref{constraints_courant_algebroid}) (see discussion in the
Appendix \ref{red_courant_algbd}) and $\theta_H$ is invariant under the action of the corresponding hamiltonian vector fields.
The symplectic reduction still gives $\M_0//\C=T^*[2]T^*[1](M/G)$ and the hamiltonian $\theta_H$ descends to $\theta_H^{red}$
that defines the quotient Courant algebroid. The $BFV_2$ charge is easily seen to be
\begin{equation}
\label{BFV_charge_courant_reduction}
\Theta_H = \theta_H+\Theta_1   \;,
\end{equation}
where $\Theta_1$ is the same in (\ref{BFV2charge}) provided on takes $\mu^{(1)}_{G\lambda}$ as degree one constraints.
This is the BFV-model for the Courant algebroid reduction. Finally, the new hamiltonian $\Theta_H$ defines another 
Courant algebroid structure on the pseudo euclidean vector bundle $E=(T+T^*)(M\times\g^*)$.

\subsection{Courant algebroid related to a Poisson action}\label{subsec: BFV_2 for poisson actions}
We can further modify the above construction by taking into account the subalgebra of the algebra of constraints given
by those of degree two, {\it i.e.} let us consider the coisotropic
$\C'=\mu^{(2)}_G{}^{-1}(0) \subset \M_0$. In this
case, $\M_0// \C' = T^*[2](\frac{T^*[1]M}{G})$ and the $BFV_{2}$
construction yields
\begin{equation*}
\M'= T^*[2](T^{\ast }[1]M\times \g[1])
\end{equation*}%
endowed with canonical degree $2$ symplectic structure $\omega_{2}'$
and with the $BFV_{2}$ charge 
\begin{equation}\label{BFV2_charge_zeropoisson}
\Theta_1'=(\mu^{(2)}_G)_a\xi ^{a}+\frac{1}{2}%
f_{bc}^{a}p_{\xi ^{a}}\xi ^{b}\xi ^{c}
\end{equation}%
where $\xi ^{a}\in C^{1}(\mathfrak{g}[1])$ and $p_{\xi ^{a}}\in C^{1}(%
\mathfrak{g}^{\ast }[1])$ are conjugate coordinates on $T^{\ast }[2](%
\mathfrak{g}[1])$.

Now we want to bring a Poisson structure $\pi=\pi^{ij}b_ib_j $ on $M$ into the picture.
To that end, let us consider the hamiltonian $\theta _{\pi }\in
C^{3}(T^{\ast }[2]T^{\ast }[1]M)$%
\begin{equation*}
\theta _{\pi }=p_{b_{j}}p_{x^{j}}+p_{x^{i}}\pi ^{ij}b_{j}+\frac{1}{2}%
p_{b_{i}}\partial _{i}\pi ^{kl}b_{k}b_{l}
\end{equation*}%
satisfying%
\begin{equation*}
\{\theta _{\pi },\theta _{\pi }\}=0\;.
\end{equation*}%
The hamiltonian $%
\theta _{\pi }$ corresponds to the Courant algebroid structure on $T^{\ast
}M\oplus TM$ obtained as the \emph{double} of the Lie bialgebroid $(T_{\pi
}^{\ast }M,TM)$ \cite{LWZ}, which in turn is isomorphic (though not equal) to
the standard Courant algebroid structure (see also the Lemma below).

If the $G$ action is Poisson,  {\it i.e.} it satisfies (\ref{eq: PL action and pi}), then one can check 
that $\theta_{\pi }$ is reducible, {\it i.e.} it satisfies 
$$
\{\theta_\pi, (\mu_G^{(2)})_a\} = \tilde{f}_a^{bc} (\mu_G^{(1)})_b (\mu_G^{(2)})_c  ~~~.
$$

By direct check we obtain the following characterization of the $BFV_2$ charge.

\smallskip
\begin{lemma}
The $BFV_2$ charge is given by
\begin{equation}
\Theta_\pi=\Theta_{\pi0}+\Theta_{\pi1}=\theta_{\pi }+(\mu^{(2)}_G)_a\xi ^{a}+\frac{1}{2}%
f_{bc}^{a}p_{\xi ^{a}}\xi ^{b}\xi ^{c}
-\tilde{f}%
_{a}^{bc} (\mu_G^{(1)})_bp_{\xi ^{c}}\xi ^{a}-\frac{1}{2}\tilde{f}%
_{a}^{bc}p_{\xi ^{b}}p_{\xi ^{c}}\xi ^{a}  \label{eq: BFV_2 charge} ~~~~~~~
\end{equation}%
The symplectomorphism
\begin{equation*}
\phi_\pi :T^{\ast }[2](T^{\ast }[1]M\times \mathfrak{g}[1])\longrightarrow
T^{\ast }[2](T^{\ast }[1]M\times \mathfrak{g}[1])    ~~,~~~~~
\end{equation*}%
given on coordinate functions by%
\begin{eqnarray*}
\phi_\pi^{\ast }(p_{b_{i}}) &=&p_{b_{i}}+\pi ^{ij}b_{j}+v_{a}^{i}\xi ^{a} \\
\phi_\pi^{\ast }(p_{x^{i}}) &=&p_{x^{i}}+\frac{1}{2}\partial _{i}\pi
^{kl}b_{k}b_{l}-\partial _{i}v_{a}^{k}\ \xi ^{a}b_{k} \\
\phi_\pi^{\ast }(p_{\xi ^{a}}) &=&p_{\xi ^{a}}-v_{a}^{i}b_{i}
\end{eqnarray*}%
and the identity on the other coordinates, satisfies
\begin{equation*}
\Theta_{\pi}=\phi_\pi^{\ast }(\Theta^{0})~,~~
\end{equation*}
where 
\begin{equation}
\Theta^{0}:=p_{b_{i}}p_{x^{i}}+\frac{1}{2}f_{bc}^{a}p_{\xi ^{a}}\xi
^{b}\xi ^{c}-\frac{1}{2}\tilde{f}_{a}^{bc}p_{\xi ^{b}}p_{\xi ^{c}}\xi ^{a}
\label{eq: BFV_2 charge in simple coords}
\end{equation}
is the hamiltonian corresponding to the product Courant algebroid
structure of the standard
$\left( T^{\ast
}M\oplus TM\right)$ with the double of the bialgebra $(\mathfrak{g}\oplus \mathfrak{g}^{\ast })$.
\end{lemma}

The main point here is that the submanifold $\mathcal{L}_\pi \subset
\M'$ of the $BFV_2$ construction above obtained by setting $\phi ^{\ast }(p_{b_{i}})=\phi
^{\ast }(p_{x^{i}})=\phi ^{\ast }(p_{\xi ^{a}})=0$, is lagrangian and corresponds
to the Dirac structure
of $\left( T^{\ast }M\oplus TM\right) \times (\mathfrak{g}\oplus \mathfrak{g}%
^{\ast })$ which encodes the Poisson-Lie action of $G$ on $(M_0,\pi )$ as
described in \cite{BCS}. 

Finally, we mention that the associated 3d AKSZ-BFV models will be the
Courant sigma models with target the Courant algebroid $\left( T^{\ast }M\oplus TM\right) \times (\mathfrak{g}\oplus \mathfrak{g}%
^{\ast })$. The properties of these models with respect to reduction
should go along the general lines of Section \ref{general_BFV_AKSZ},
and will be studied elsewhere.

\begin{remark}
One could also discuss the
corresponding 3d AKSZ-BFV models in which $\Sigma_3$ has boundary and
the boundary condition for the fields is taken to be the one determined by $\mathcal{L}_\pi$.
\end{remark}

\bigskip
\bigskip

\section{Summary and outlook}
\label{conclusions}

We have presented a framework to encode systematically the reduction
of the target space geometry into ASKZ sigma models. This is provided
by the construction
of the $BFV-ASKZ$ model where one has a clear picture of the role of the
ingredients involved. We then studied how this model is related to the underlying
reduced one, showing that the
correlators of the reduced theory can be formally obtained from the $BFV-ASKZ$ model by
imposing suitable gauge fixings. Moreover, we implemented our
construction in particular cases and showed that we recover several of the
gauged models that were previously considered in the literature, thus
providing a clear conceptual context for their study.

Notice, however, that the framework that we discussed is very general and can be applied to all kind of symplectic reduction in graded symplectic geometry.
%Provided the constraints are sufficiently regular, homological perturbation theory assures the existence of the $BFV$ hamiltonian and so the whole 
%construction of the $BFV-AKSZ$ model. Nevertheless, the explicit construction of the hamiltonian can be cumbersome in general. The examples that we 
%considered in this paper are always related to a group action where this problem can be treated much more easily. There are several interesting 
%examples that could be addressed and the problem of explicitly
%computing the $BFV$ hamiltonian must be solved. 
So, let us now discuss some possible future directions.

Consider a Poisson manifold obtained as a quotient of a symplectic one; in this case
the $BFV-AKSZ$ theory will give an extension of the PSM with target a symplectic manifold by adding ghost and antighosts. In the symplectic case the PSM is
equivalent to the A-model by choosing the gauge fixing given by the choice of an almost complex structure (see \cite{BZ 08}); putting together this with the ghost
gauge fixing we can think of studying the non perturbative properties
of the reduced model, so far unaccessible for the generic Poisson case. 

On the other hand, when a Lie group $G$ acts by symmetries on a
Poisson manifold, the perturbative quantization of the corresponding $BFV-AKSZ$ model
could be related to the $G-$\emph{equivariant} version of Kontsevich's
formality. Something similar could be also explored for the case in which the
$G$ action is hamiltonian and one wants to get star products on $\mu_0^{-1}(0)/G$. 

More generally, take an integrable Poisson manifold with symplectic groupoid $\G(M)$. Then the right invariant vector fields define a 
coisotropic submanifold of $T^*[1]\G$ whose symplectic reduction is $T^*[1]M$. The $BFV$ manifold will be a non negatively graded $n=1$ 
symplectic manifold, {\it i.e.} a Poisson manifold containing the symplectig groupoid as a submanifold. Of course in this case the problem of
explicitly computing the $BFV$ hamiltonian can be rather difficult to
be solved in general. 

We will come back to these examples in the future.

%%%%%%%%%%%%%%%%%%%%%%%%%%%%%%%%%%%%%%%%%%%%%%%%%%%%%%%%%%%%%%%%%%%%%%%%%%%%%%%%%%%%%%%%%%%%%%%%%%%%%%%%%%%%%%%%%%%%%%%%%%%%%%%%%%%%%%%%%%%%%%%%%%%%%%%%%%%%%%%%%%%%%%

%%%%%%%%%%%%%%%%%%%%%%%%%%%%%%%%%%%%%%%%%%%%%%%%%%%%%%%%%%%%%%%%%%%%%%%%%%%%%%%%%%%%%%%%%%%%%%%%%%%%%%%%%%%%%%%%%%%%%%%%%%%%%%%%%%%%%%%%%%%%%%%%%%%%%%%%%%%%%%%%
\appendix

\section{Quantization of Hamiltonian systems with symmetry}

\label{app: ham syst}

Let $(T^{\ast }Q,\omega _{0},G,J,H\in C^{\infty }(T^{\ast }Q))$ be a
hamiltonian system with symmetry, where the moment
map is $J:T^{\ast }Q\longrightarrow \mathfrak{g}^{\ast }$. The
underlying first class constrained classical action is 
\begin{equation*}
S_{cl}=\int\limits_{I}dt \ p_i\dot{x}^{i}-H(x,p)  - \lambda^a J_{a}(x^i,p_i)
= S_{cl}^0 -\int\limits_{I}dt \lambda^a J_{a}(x^i,p_i)
\end{equation*}%
where $I=[0,T]$ denotes
a time interval.
Assuming $0$ is a
regular value for $J$, the corresponding $%
BFV_{n=0}$ construction yields $(\mathcal{M}:=T^{\ast }Q\times \mathfrak{g}%
[1]\times \mathfrak{g}^{\ast }[-1],\omega _{0}\oplus \varpi _{0},\Theta
_{1}) $ as in Section \ref{BFVconstruction}. The BFV charge is given
by
$$
\Theta_1(x^i,p_i,\xi^a,p_a)= \xi^a J_a(x^i,p_i) - \frac{1}{2} f^c_{ab}
p_c \xi^a \xi^b
$$
where $x^i,p_i$ are canonical coordinates on $T^*Q$, $\xi^a$ are deg
$1$ coordinates on $g[1]$ and $p_a$ their deg $-1$ conjugates.

We shall mimic the BFV-AKSZ procedure of Section \ref{general_BFV_AKSZ}, but adapted to the
non-topological case as follows. Consider the space of BV
fields and antifields to be%
\begin{equation*}
\fF=Map\{T[1]I, \mathcal{M}\} \simeq T[-1]T^*PQ
\end{equation*}%
where $PQ=\{I \to Q\}$
denotes the path space and the fields are assumed to satisfy
appropriate boundary conditions. We take the extended BV action to be
\begin{equation*}
S_{BV}=\int\limits_{T[1]I}\boldsymbol{p}_{x^{i}}d\boldsymbol{x}^{i}+\boldsymbol{p}%
_{\xi ^{a}}d\boldsymbol{\xi }^{a} - \Theta_1 (\boldsymbol{p}%
_i,\boldsymbol{x}^{i},\boldsymbol{\xi
}^{a},\boldsymbol{p}_{c})-H(x,p_{x})\theta _{t} \in C(\fF)
\end{equation*}%
Notice that by setting the BV fields with non-zero degree to zero one
recovers the classical constrained action $S_{cl}$. 
The only difference with the usual AKSZ construction is the last
term involving $\theta _{t}\equiv dt$ on $T[1]I$. 
The CME is still satisfied%
\begin{equation*}
\{S_{BV},S_{BV}\}=0
\end{equation*}%
but it only reflects the $G-$symmetry, the $Diff(I)$ invariance is
broken by the term with $H\neq 0$. This agrees with the fact that
the theory is non-topological since it has non-trivial hamiltonian dynamics.

%%%%%%%%%%%%%%%%%%
Let us now consider an appropriate gauge fixing. For simplicity, assume
$G$ acts on $Q$ freely and properly and that the action on $T^*Q$ is the lifted one.
Let us choose $\Gamma \subset Q$
transverse to the $G$-action and set $W: = T^*Q|_\Gamma \subset
T^*Q$. Notice these are chosen so that $W \cap J^{-1}(0) \simeq
T^*(Q/G)$ yields the reduced phase space.

% Now, by using Legendre transformation one gets $T[-1]T^*PQ  \simeq T^*[-1]T^*PQ$. 
Noticing that $PW \subset
PT^*Q\simeq T^*PQ$, we can thus define the Lagrangian
gauge fixing to be $\fL := N^*[-1](PW) \subset T^*[-1]T^*PQ\simeq
\fF$.

\begin{remark}
  Let $\tilde{x}^\alpha, y^a$ denote adapted coordinates for which $Q \simeq
  Q/G \times G$, and $\tilde{p}_\alpha, z_a$ their conjugates on
  $T^*Q$. Then, $PW \simeq \{ y^a(t) =0 \}$ and
$$
\fL \simeq \{ y^a(t)=0, \tilde{x}^\alpha_{(1)}(t)=0=
  \tilde{p}_\alpha^{(1)}(t), y^a_{(1)}(t)=0 \}
$$
recalling the notation for the BV fields is $ev^*\phi = \phi (t) +
\phi_{(1)}(t) \theta_t$.
\end{remark}

One can then compute the partition function as follows
\begin{eqnarray*}
Z = \int\limits_\fL e^{iS_{BV}} & = &\int\limits_{PW} e^{iS_{cl}^0|_{PW}} \int D\xi^a
D\xi^a_{(1)} Dz_a^{(1)} e^{-i\int_I J_a \xi^a_{(1)} + \xi^a \{ J_a,
  y^b\} z_b^{(1)} } \\
 &=& \int\limits_{T^*PQ} e^{iS_{cl}^0} \delta(J_a) \delta(y^b) det(\{J_a,y^b\})
 \\
&=& \int\limits_{T^*P(Q/G)} e^{i\tilde{S}_{cl}^0} det(\{J_a,y^b\}|_{J_a=0=y^b})
= Z_{red}
\end{eqnarray*}
Renaming the degree $0$ coordinates
$\xi^a_{(1)}=:\lambda^a$, the above is exactly the expression given in
\cite{Fad} for the path integral on reduced space\footnote{One can check
  that the above expression is indeed independent of the choice of adapted coordinates.}. 
See also \cite{GD99} where they also explain the relation to the usual
BFV hamiltonian quantization \cite{Hen 85}.

%%%%%%%%%%%%%%%%%%%%%%%%%%%%%%%%%%%%%%%%%%%%%%%%%%%%%%%%%%%%%%%%%%%%%%%%%%%%%%%%%%%%%%%%%%%%%%%%%%%%%%%%%%%%%%%%%%%%%%%%%%%%%%%%%%%%%%%%%%%%%%%%%%%%%%%%%%%%%%%%

\section{Basic facts of Poisson actions}

\label{subsec: Basics Poisson actions}
Let us review here basic facts of Poisson reduction. For a systematic treatment in terms of the 
supergeometric language see \cite{CZ}.

Let $G$ be a Lie group acting freely and properly
on $M$. Let $\{x^{i}\}$ be coordinates on $M_0$ and $\{b_{i}\}$ the odd
coordinates on the fibre of $T^{\ast }[1]M$. Let us introduce a basis $%
\{T_{a}\}$ of $\g\equiv \mathrm{Lie}G$ and let $v_{a}=v_{a}^{i}\partial _{i}$
be the fundamental vector field of $T_{a}\in \g$ and $(\mu^{(1)}
_{G})_{a}=v_{a}^{i}b_{i}$ the corresponding degree one momentum map which
lifts the $G$ action to an hamiltonian action on $T^{\ast }[1]M$ so that
\begin{equation*}
T^{\ast }[1](M/G)=(\mu^{(1)}_{G})^{-1}(0)/G\;.
\end{equation*}%
As usual, the above quotient means that $(\mu_{G}^{(1)})^{-1}(0)$ is the
submanifold of $T^{\ast }[1]M$ defined by the ideal $I_{\mu^{(1)}_G}\subset C^{\infty
}(T^{\ast }[1]M)$ generated by $\mu^{(1)}_{G}$ and $T^{\ast }[1](M/G)$ is the
graded manifold whose functions are $C^{\infty }((\mu_{G}^{(1)})^{-1}(0))^{inv}$,
the invariant functions on $(\mu_{G}^{(1)})^{-1}(0)$ with respect to the $G$
action. For any subgroup $H\subset G$ with Lie algebra $\h$, we denote the
moment map with $\mu_{H}^{(1)}=\mathrm{pr}_{h}\circ \mu^{(1)}_{G}$, where $\mathrm{pr}%
_{h}:\g^{\ast }\rightarrow \g^{\ast }/\h^{\perp }=\h^{\ast }$.

As in section \ref{subsec: BFV_1 for G acts}, let us assume now that $M$ is
a Poisson manifold with tensor $\pi_{M}=\frac{1}{2}\pi ^{ij}b_{i}b_{j}$, $G$
is a Poisson Lie group with tensor $\pi_{G}$ and that the action is
Poisson. This means that
\begin{equation}
\{(\mu^{(1)}_{G})_{a},\pi_{M}\}=\frac{1}{2}\tilde{f}_{a}^{bc}(\mu^{(1)}_{G})_{b}(\mu
_{G}^{(1)})_{c}\;,  \label{PLaction}
\end{equation}%
where $\tilde{f}_{a}^{bc}$ are the structure constants of $\g^{\ast }$. To
these data we can associate the following constructions (see for instance
\cite{L}):

$i$) \textit{Poisson reduction}. Let $H\subset G$ be a closed subgroup of $G$
which is coisotropic with respect to $\pi _{G}$, \textit{i.e.} $\pi
_{G}(N^{\ast }H)\subset TH$. We recall that, then, the annihilator $\h^{\perp }$ is a subalgebra of
$\g^{\ast }$; it follows from (\ref{PLaction}) that $\pi_{M}\in C^{2}(T^{\ast }[1]M)$ descends to 
$\pi_{M/H}\in C^{2}(T^{\ast }[1](M/H))$%
, \textit{i.e.} on $M/H$ there exists a unique Poisson structure such that
the projection $M\rightarrow M/H$ is Poisson.

$ii$) \textit{A Poisson structure on $M\times G^{\ast }$}. Let $G^{\ast }$
be the dual Poisson Lie group of $G$ with Poisson tensor $\pi_{G^{\ast }}$.
Then we can define on $M\times G^{\ast }$ a Poisson structure. Let $%
\{\lambda _{a}\}$ be coordinates on $G^{\ast }$ and $\{\beta ^{a}\}$ on the
fibre of $T^{\ast }[1]G^{\ast }$. Let us denote with $k^{b}=k_{a}^{b}\frac{%
\partial }{\partial \lambda _{a}}$ the left invariant vector fields of $%
G^{\ast }$. The tensor
\begin{equation}
\pi_{M\vartriangleleft G^{\ast }}=\pi _{M}+\pi _{G^{\ast
}}+v_{b}^{i}k_{a}^{b}b_{i}\beta ^{a}=\pi _{M}+\pi _{G^{\ast }}+(\mu_{G}^{(1)})_{b}k^{b}  \label{semi_direct_prod}
\end{equation}%
is Poisson as a consequence of (\ref{PLaction}).

\section{Reduction of Courant algebroids}
\label{red_courant_algbd}

The main source for reduction of Courant algebroids is \cite{BCG}. The description in terms of
graded manifold language can be found in \cite{BCMZ}. Here we will consider the particular situation of a trivially extended isotropic action.

An exact Courant algebroid is given by the following operations on the space of section of $E=TM+T^*M$:
\begin{itemize}
 \item[$i$)] the pairing $\langle v+\omega,w+\nu\rangle = \frac{1}{2}(\omega(w)+\nu(v))$\;; 
 \item[$ii$)] the Courant bracket $[v+\omega,w+\nu]=[v,w]+{\cal L}_v\nu - \iota_w\omega + \iota_w\iota_v H$
\end{itemize}
where $v,w\in {\cal X}(M)$, $\omega,\nu\in\Omega^1(M)$ and $H\in \Omega^3(M)$ is a closed three form representing the so called {\it Severa} class of $E$.
It can be encoded in the graded manifold language as follows. Let us consider the symplectic graded manifold $T^*[2]T^*[1]M$
of degree $2$. In terms of the local Darboux coordinates $(x^i,b_i,p_{b_i},p_{x^i})$ of degree $(0,1,1,2)$, the homological hamiltonian 
encoding the pairing and the Courant bracket reads
$$\theta_H=p_{x^i}p_{b_i}+\frac{1}{6} H_{ijk}p_{b_i}p_{b_j}p_{b_k}~~~~~.$$

Let the Lie group $G$ act on $M$ freely and let $v_X$ denote
the fundamental vector field of $X\in\g$. Following \cite{BCG}, a {\it trivial extension} of this action to the Courant algebroid $E$ is a map 
$\rho:\g\rightarrow \Gamma(E)$  preserving the splitting of $E$ and the Courant bracket.
If we denote $\rho(X)=v_X+\lambda_X \in TM+T^*M$ for $X\in\g$, these two conditions mean that for any $X,Y\in\g$ we have
$$v_{[X,Y]}=[v_X,v_Y]~,~~~  \iota_{v_X}H=d\lambda_X~~,~~~~~~ \lambda_{[X,Y]}={\cal L}_{X}\lambda_Y~~.$$
As a consequence
$\Phi=H+\sum_a t_a X_a\in C^3(T[1]M\times \g^*[2])$ is an equivariant form in the Cartan complex for equivariant cohomology and one computes
$d_G(\Phi)=\langle \rho(\g),\rho(\g)\rangle$, where $d_G=d + \sum_a t_a \iota_{v_a}$.
The action is {\it isotropic} if $d_G(\Phi)=0$ so that $\Phi$ is an equivariant extension of $H$; we say that $\rho$ defines an isotropic trivial 
extension of $v$.

The general procedure of reduction described in Proposition 3.6 of \cite{BCG} gives a reduced Courant algebroid $E_{red}$ on $M/G$ which is exact.
The reduced Severa class is $[\Phi]\in H^3_G(M,\R)\sim H^3(M/G)$.

In terms of graded manifold language, the above setting is described by the constraints 
$(\mu^{(2)}_G)_a\in C^1(T^*[2]T^*[1]M)$ and $(\mu^{(1)}_{G\lambda})_a\in C^2(T^*[2]T^*[1]M)$ defined as
\begin{equation}\label{constraints}
(\mu^{(2)}_G)_a= v_a^i p_i - \partial_iv_a^j b_j p_{b_i} \:,~~~~~ (\mu^{(1)}_{G\lambda})_a = v_a^i b_i + \lambda_{ai}p_{b_i}\;.
\end{equation}
Since the action is isotropic, they satisfy the algebra
\begin{equation}\label{constraints_courant_algebroid}
\{(\mu^{(2)}_G)_a,(\mu^{(2)}_G)_b\}= f_{ab}^c (\mu_G^{(2)})_c ~~,~~~~~ \{(\mu_G^{(2)})_a,(\mu^{(1)}_{G\lambda})_a\}= f_{ab}^c (\mu_{G\lambda}^{(1)})_c~,~~~~
\{(\mu_{G\lambda}^{(1)})_a,(\mu^{(1)}_{G\lambda})_a\}=0 \;.
\end{equation}

Moreover, since $H$ is $\g$-invariant then we have that
$$
\{(\mu^{(2)}_G)_a,\theta_H\} = \{(\mu^{(1)}_{G\lambda})_a,\theta_H\}= 0\;,
$$
showing that $\theta_H$ is reducible. 

\section{Relation to Poisson equivariant cohomology}

\label{App: M times G* and Poisson equivar cohomology}
Equivariant Poisson cohomology  is defined in \cite{Gi} for a Poisson Lie $G-$manifold $%
M$ endowed with a (pre-)momentum map.
Here we will concentrate on the case in which $(G,\pi_G=0)$ acts hamiltonianly on $(M,\pi_M)$ with
equivariant moment map $\mu^{(0)}_G:M \rightarrow
\mathfrak{g}^*$. This corresponds to the case $G^*=\mathfrak{g}^*$ and the $BFV_1$ construction has been discussed at the end of Subsection 
\ref{subsec: BFV_1 for G acts}.

Let us first say a few words about equivariant Poisson cohomology of $(M,\pi_M,G,\mu^{(0)}_G)$ following \cite{Gi}. One considers a Weil model complex as
\begin{equation*}
 A:=(\mathfrak{X}(M) \otimes W(\mathfrak{g}), d_A = d_{\pi_M} + d_W)
\end{equation*}
where
$W(\mathfrak{g})=S\mathfrak{g}^* \otimes \Lambda \mathfrak{g}^*$
denotes the Weil algebra of $\mathfrak{g}$ endowed with the usual Weil
differential $d_W$ and $d_\pi=[\pi, \ ]$  denotes Poisson cohomology
differential. Next, define the subcomplex $A_b \subset A$ formed by basic elements $F \in A$ such that
\begin{eqnarray*}
 (i_{d(\mu^{(0)}_G)_a}+I^W_a)F=0 \\
(L_{v_a}+L^W_a)F=0
\end{eqnarray*}
where $i_{d(\mu^{(0)}_G)_a}$ denotes contraction and $ L_{v_a}$ Lie
derivative on $\mathfrak{X}(M)$ while $I^W_a$, $L^W_a$ denote the usual contraction and $\mathfrak{g}$-action operations on $W(\mathfrak{g})$. 
Equivariant Poisson cohomology $H_G(M,\pi_M,\mu_G^{(0)})$ is defined as the cohomology of the subcomplex $(A_b, d_A|_{A_b})\subset (A,d_A)$.

The corresponding $BFV_1$ space has been studied at the end of subsection \ref{subsec: BFV_1 for G acts}. As a degree one symplectic manifold it is
\begin{eqnarray*}
 \M=T^{\ast }[1](M \times \mathfrak{g}[1]\times \mathfrak{g}[2])
\end{eqnarray*}
and the $BFV_1$-charge $\Theta$ is given in (\ref{eq: BFV_1 charge for moment map g*}). Notice that 
$C(\M)=\mathfrak{X}(M) \otimes W(\mathfrak{g}) \otimes
C(\mathfrak{g}^*[0] \times \mathfrak{g}^*[-1])$. For the BFV differential
$Q=\{ \Theta, \ \}$, we get in particular
\begin{eqnarray*}
 Q(\xi^a_{(1)})&=&\xi^a_{(0)} - \frac{1}{2} f^a_{bc} \xi^b_{(1)} \xi^c_{(1)} \\
Q(\xi^a_{(0)})&=&- f^a_{bc} \xi^b_{(1)} \xi^c_{(0)} \\
Q(\alpha)&=& d_{\pi_M}\alpha +  \xi^a_{(1)} L_{v_a}\alpha - \xi^a_{(0)} i_{d(\mu^{(0)}_G)_a}\alpha
\end{eqnarray*}
where $\alpha \in \mathfrak{X}(M)$.
Thus $B:=\mathfrak{X}(M) \otimes W(\mathfrak{g})$ is a sub
DGA of $(C(\M),Q)$ and let us denote by $d_K = Q|_B$ the induced
differential. We recognize the Weil differential $d_W$ as the restriction 
$d_K|_{W(\mathfrak{g})}$. Moreover, the differential $d_K$ is analogous to the one defined by Kalkman (\cite{Kalkman}) for 
his BRST model of equivariant de Rham cohomology.

We can thus follow \cite{Kalkman} further to get an isomorphism $\psi : B \rightarrow A$ of DGA's defined by 
$\psi=exp(-\xi^a_{(1)} i_{d(\mu^{(0)}_G)_a})$ so that we get a chain of inclusions
\begin{equation} \label{eq: inclusion Pequiv in BFV}
 (A_b,d_{A_b}) \hookrightarrow (A,d_A) \simeq^\psi (B,d_K) \hookrightarrow (C(\M), Q)
\end{equation}
Moreover, analogously to \cite{Kalkman},
$\psi^{-1}(A_b)=(\mathfrak{X}(M)\otimes S(\mathfrak{g}^*))^G$ gives
the Cartan model for Poisson equivariant cohomology (\cite{Gi}). 

Finally, as observed in Subsection \ref{subsec: BFV_1 for G acts}, the Poisson-Weil model of
\cite{Zuc07,Zuc08} is the corresponding BFV-AKSZ model so that the above construction gives 
a clear explanation of the relation between one sector of the observables of the Poisson-Weil 
model with Poisson equivariant cohomology as observed in \cite{Zuc08}. 

\begin{remark}
  In the case where we deal with a general Poisson-Lie action with
  moment map $\mu_G^{(0)}:M \rightarrow G^*$, one would need to consider
  a non-linear version of the $BFV_{1}$ construction as in Section \ref{subsec: BFV_1 for G acts}.
\end{remark}
\section{The zero modes case of AKSZ commuting with BFV}
\label{app: zero modes AKSZ commutes BFV}
In this section we give the details of the proof of the
statement of Section \ref{general_BFV_AKSZ} about AKSZ commuting with BFV in the context of finite
dimensional 'zero modes' AKSZ theories (see details in \cite{BMZ}).

\bigskip

Let $X_\Sigma$ be the sheaf over a point corresponding to the
cohomology ring $H_{dR}(\Sigma_{n+1})$ of the source and $(M_n,\omega
_{M_{n}},\theta_{M_{n}},C_{n})$ a degree $n$ symplectic manifold
endowed with reduction data $C_n$. Assume that $\Sigma_{n+1}$ is
compact without boundary. The zero modes field space is
\begin{equation*}
 \fF^Z_0= Map\{ X_\Sigma \rightarrow M_{n} \}
\end{equation*}
and the evaluation map is denoted by $ev_0:\fF^Z_0\times X_\Sigma
\rightarrow M_{n}$. From integration over $\Sigma_{n+1}$ we get an
integration $\int ds$ defined by $ \int ds \ s =1$
for $s \in H^{n+1}_{dR}(\Sigma_{n+1})$ being a chosen normalized
nontrivial top form. Using these ingredients, one can endow $\fF^Z_0$
with a degree $-1$ symplectic structure $\omega_{\fF^Z_0}$ via an
analogous formula to the one on
$\fF_0$ as described in Section \ref{Section_AKSZ}. Choosing a basis $\alpha_{(l),i_l}$ for $H^l_{dR}(\Sigma_{n+1})$,
we get for any function on the target $f \in C(M_n)$
\begin{equation*}
 ev_0^*f=\sum_{l,i_l} f^{(l),i_l} \ \alpha_{(l),i_l}
\end{equation*}
where $f^{(l),i_l} \in C(\fF^Z_0)$. One can then show by direct computation
the following:

\begin{lemma} \label{lma: zero modes}
 Assume $C_n$ is regular and given as a level set $C_n=\{ \mu_a =0 \} \subset M_n$, then
 \begin{itemize}
  \item The submanifold $\fC_\mu^Z:= Map\{ X_\Sigma \rightarrow C_n \} \subset \fF^Z_0$ is given by the vanishing of all components
  \begin{equation}
  \mu_a^{(l),i_l} =0
  \end{equation}
  of $ev_0^*\mu_a$.
  \item $\fC_\mu^Z$ is coisotropic inside of $\fF^Z_0$, namely,
  \begin{equation*}
  \{ \mu_a^{(l),i_l}, \mu_b^{(m),j_m} \}_{\fF^Z_0} \in I\langle \mu_c^{(k),i_k} \rangle
  \end{equation*}
\item
 when $f \in C(M_n)$ is reducible by $C_n$, i.e. $\{ \mu_a, f \}_{M_n}
 = A_{a}^c \mu_c$, then $f^Z:=\int ds \ ev_0^*f$ is reducible by $\fC_\mu^Z$:
\begin{equation*}
 \{ \mu^{(l),i_l}_a, f^Z \}_{\fF^Z_0} \propto I\langle \mu^{(k),i_k}_c\rangle
\end{equation*}
 \end{itemize}

\end{lemma}

%The proof can be done using the statements of Appendix \ref{App: computing brackets}

In particular, the zero modes BV action $S_{\theta}^Z = \int ds \
ev_0^* \theta_{M_n} \in \fF^Z_0$ is reducible by $\fC_\mu^Z \subset \fF^Z_0$ since $\theta_{M_n}$ is reducible by $C_n \subset M_n$.

We thus get that $( \fF^Z_0, \omega_{\fF^Z_0}, S_{\theta}^Z,
\fC_\mu^Z)$ defines reduction data on $\fF^Z_0$. A $BFV_{-1}$ model
for this reduction data will be thus given by a degree $-1$ symplectic manifold that looks like
\begin{equation} \label{eq: BFV(AKSZ) space}
 BFV_{-1} \approx \fF^Z_0\times \{ \eta^a_{(l),i_l} \} \times  \{ p_{\eta^a}^{(l),i_l} \}
  \end{equation}
where we have one \emph{ghost} $\eta^a_{(l),i_l}$ for each constraint
$\mu_a^{(l),i_l} \in C(\fF^Z_0)$ while the $p_{\eta^a}^{(l),i_l}$
denote their conjugate antighosts.
The corresponding $BFV_{-1}$ charge will take the form
\begin{equation} \label{eq: BFVoAKSZ charge}
 \Theta^Z := \sum_{l,i_l} \mu_a^{(l),i_l} \eta^a_{(l),i_l} + S_{\theta}^Z + O(p_\eta)
  \end{equation}
Notice that $\Theta^Z$ must have degree 0 as $S_{\theta}^Z$ does and that, then, the degree of the ghosts $\eta^a_{(l),i_l}$ must be $(l-k_a)$ where $k_a$ is the degree of $\mu_a \in C(M_n)$.

With this set up, we can then state that AKSZ commutes with BFV:

\begin{proposition}
 Let $(\M,\Omega,\Theta)$ be a $BFV_n$ model for the target reduction
 data $(M_n, \omega_n, \theta_n,C_n)$. Then, the construction $(\fF^Z,\omega_{\fF^Z},\Theta^Z)
 :=AKSZ(X_\Sigma,(\M,\Omega,\Theta))$ gives a $BFV_{-1}$ model for the
 reduction data 
$(\fF^Z_0, \omega_{\fF^Z_0}, S_{\theta}^Z, \fC_\mu^Z)$.
\end{proposition}
{\it Proof}:
Let us consider the identification
$$
 \eta^a_{(l)} = \xi^{a,(m)}
$$
for $m=n+1-l$, with $ev_0^*\xi^a=\sum_{m,i_m} \xi^{a, (m),i_m} \alpha_{(m),i_m}$ and $\xi^a$ being the ghost in $\M$ associated to the constraint $\mu_a$ on the target $M_n$.
Then, one can direclty check that by applying the AKSZ construction, both the space $\fF^Z=Map\{ X_\Sigma \to \M \}$ and
the charge $\Theta^Z = \int ds \ ev_0^*
\Theta$ have the desired form $( \ref{eq:
  BFV(AKSZ) space} )$ and $( \ref{eq: BFVoAKSZ charge} )$,
respectively. Moreover, by construction, $\Theta^Z$ is homological, $\{
\Theta^Z,\Theta^Z \}=0 $. $\square$

\begin{remark}
Recall from \cite{BMZ}, that the space of zero modes $\fF_0^Z$ can be
obtained from the full AKSZ field space $\mathfrak{F}_0=\{
T[1]\Sigma_{n+1} \rightarrow M_n \}$ via reduction  $\fF_0^Z=\fF_0 // \mathfrak{Z}$ by the coisotropic
$\mathfrak{Z} = \{ \Phi \in \fF_0 :D\Phi=0 \}$.
One can check that reduction by $\mathfrak{Z}$ is formaly compatible with reduction by $\mathfrak{C}_0:=
\{ T[1]\Sigma_{n+1} \rightarrow C_n \}$. In fact, one can also obtain
the computations of Lemma \ref{lma: zero modes} directly as reduction by $\fC_0$ of
formal computations on $\mathfrak{Z} \subset \fF_0$. In particular, imposing
$D\Phi=0$ brings reduction data $(\fF_0,
\omega_{\fF_0},S_{\theta},\fC_0)$ to the zero modes reduction data $( \fF^Z_0, \omega_{\fF^Z_0}, S_{\theta}^Z, \fC_\mu^Z)$.
\end{remark}

%%%%%%%%%%%%%%%%%%%%%%%%%%%%%%%%%%%%%%%%%%%%%%%%%%%%%%%%%%%%%%%%%%%%%%%%%%%%%%%%%%%%%%%
%%%%%%%%%%%%%%%%%%%%%%%%%%%%%%%%%%%%%%%%%%%%%%%%%%%%%%%%%%%%%%%%%%%%%%%%%%%%%%%%%%%%%%%%%%%%%%%%%%%%%%%%%%%%%%%%%%%%%%%%%%%%%%%%%%%%%%%%%%%%%%%%%%%%%%%%%%%%%%%%


\begin{thebibliography}{99}
%\bibitem{AM} Abraham R. and Marsden J.: \emph{Foundations of Mechanics}.
%Second Edition, Addison-Wesley, Menlo Park, California.

\bibitem{Alexandrov:1995kv} M.~Alexandrov, M.~Kontsevich, A.~Schwartz and
O.~Zaboronsky, ``The Geometry of the master equation and topological quantum
field theory,'' Int.\ J.\ Mod.\ Phys.\ A \textbf{12} (1997) 1405
[arXiv:hep-th/9502010]. %%CITATION = HEP-TH 9502010;%%
%\cite{Batalin:1981jr}

\bibitem{BFV} I. Batalin, E. Fradkin, ``A generalized canonical formalism and quantization
of reducible gauge theories'', Phys. Lett. {\bf 122B} (1983), 157.

\bibitem{BV1} I. Batalin and G Vilkovisky, ``Gauge algebra and
  quantization'', Phys. Lett., {\bf 102B} (1981) 27.

\bibitem{BV2} I. Batalin, and G. Vilkovisky, ``Quantization of gauge
  theories with linearly dependent generators'', 
Phys. Rev. {\bf D29} (1983), 2567.

%\cite{Bonechi:2011um}
\bibitem{Bonechi:2011um}
  F.~Bonechi, A.~S.~Cattaneo and P.~Mn\"ev,
  ``The Poisson sigma model on closed surfaces,'' J.High Energy Phys. {\bf 2012} JHEP01(2012)099
  [arXiv:1110.4850 [hep-th]].
  %%CITATION = ARXIV:1110.4850;%%

%\bibitem{BZ 05} F. Bonechi and M. Zabzine, ``Lie algebroids, Lie Groupoids and TFT''.
%J.Geom.Phys. {\bf 57} 3, 731-744. [arXiv:math.SG/0512245]

\bibitem{BZ 08} F. Bonechi and M. Zabzine, ``Poisson sigma
model on the sphere'', Commun.\ Math.\ Phys.\  {\bf{285}} (2009) 1033
[arXiv:hep-th/0706.3164].

\bibitem{BMZ} F. Bonechi, P. Mn\"ev and M. Zabzine, ``Finite dimensional AKSZ-BV theories,''
   Lett.\ Math.\ Phys.\  {\bf 94} (2010) 197
  [arXiv:0903.0995 [hep-th]].

%\bibitem{BC07} H. Bursztyn and M.Crainic, 
%``Dirac geometry, quasi-Poisson
%actions and $D/G$-valued moment maps,'' 
%J. Differential Geom. {\bf 82} (2009), no. 3, 
%[arXiv:0710.0639 [math.DG]]

\bibitem{BCS} H. Bursztyn, M. Crainic and P. Severa,  ``Quasi-Poisson
structures as Dirac structures,''  Travaux Math\'ematiques XVI (2005), 41-52.

\bibitem{BCG} H. Bursztyn, G. R. Cavalcanti and M. Gualtieri 
 ``Reduction of Courant algebroids and generalized complex structures,'' 
  Adv. Math., {\bf 211} (2), 2007, 726-765 [math/0509640[math.DG]]

\bibitem{BCMZ} H. Bursztyn, A. Cattaneo, R. Mehta and  M. Zambon, in preparation.

\bibitem{CH} A. Cabrera and H.-C. Herbig,  ``BFV\ complex for graded
manifolds,'' in preparation.

%\bibitem{CF00} A. Cattaneo, G. Felder,: \emph{Poisson sigma models and
%symplectic groupoids, }arXiv:math/0003023v1

\bibitem{CF01} A. Cattaneo and G. Felder, ``On the AKSZ formulation of the Poisson Sigma Model''. 
Lett.Math.Phys. {\bf 56} 2 (2001) 163-179. [arXiv:math/0102108] 

\bibitem{CMR12} A. Cattaneo, P. Mnev, N. Reshetikhin, ``Classical BV
  theories on manifolds with boundaries'', [arXiv:math-ph/1201.0290]


\bibitem{CZ} A. Cattaneo and  M.Zambon, ``A supergeometric approach to Poisson reduction'', [arxiv:math/1009.0948].

\bibitem{Fad} L.~D.~Faddeev,
  ``Feynman integral for singular Lagrangians,''
  Theor.\ Math.\ Phys.\  {\bf 1}, 1 (1969)
  [Teor.\ Mat.\ Fiz.\  {\bf 1}, 3 (1969)].

\bibitem{Gi} V. L. Ginzburg, ``Equivariant Poisson
cohomology and a spectral sequence associated with a moment
map''. Internat. J. Math., 10 (1999), 977--1010 [arxiv:dg-ga/9611002].

\bibitem{GW} V.L.Ginzburg and  A.Weinstein, ``Lie Poisson structure on some Poisson Lie
groups'', J. Amer. Math. Soc. 5 (1992), 445.

\bibitem{GD99} M. Grigoriev and P.H. Damgaard,
  ``Superfield BRST Charge and the Master Action,''
   Phys.Lett. B{\bf 474} (2000) 323-330 [arXiv:hep-th/9911092].

\bibitem{Hen 85} M. Henneaux, "Hamiltonian form of the path integral for
theories with a gauge freedom", Physics Reports, {\bf 126}, 1,(1985) 1-66.

\bibitem{Hen-Tei} M.\ Henneaux  and C. Teitelboim, \emph{Quantization of gauge
systems}. Princeton University Press, Princeton, (1992).

\bibitem{Herbig} H-C. Herbig: ``Variations on Homological Reduction'',
PhD thesis  [arXiv:0708.3598[math.QA]].

\bibitem{Kalkman} J. Kalkman, \emph{A BRST\ model applied to symplectic
geometry}, PhD Thesis Utrecht, 1993 [hep-th/9308132].

\bibitem{Kimura} T. Kimura, ``Generalized classical BRST and reduction
of Poisson manifolds''. Comm Math Phys 151 (1993)\ 155-182.

%\bibitem{KS} B. Kostant and S. Sternberg, ``Symplectic reduction BRS
%cohomology and infnite dimensional Clifford algebras'', Annals of Physics 176
%(1987) 49-113.

\bibitem{LWZ}  Z.-J. Liu, A. Weinstein and P. Xu,  ``Manin triples for
Lie bialgebroids'', J. Diff. Geom. 45 (1997), 547-574 [arxiv:dg-ga/9508013].

\bibitem{losev}
A. Losev, ``BV formalism and quantum homotopical structures,'' Lectures at GAP3, Perugia, 2005.

\bibitem{L} H-J. Lu, \emph{Multiplicative and Affine Poisson Structures on Lie Groups},
 PhD thesis, Berkeley, can be downloaded from 
http://hkumath.hku.hk/$\sim$jhlu/thesis.pdf

\bibitem{LS} S.L. Lyakhovich and  A.A. Sharapov, ``BRST theory without
Hamiltonian and Lagrangian'', J.High Energy Phys. {\bf 2005}, JHEP03 (2005) [arXiv:hep-th/0411247]

\bibitem{Mn} P. Mn\"ev, ``Discrete BF theory'', PhD thesis, [arXiv:08091160].

\bibitem{Qiu:2011qr} 
  J.~Qiu and M.~Zabzine,
  ``Introduction to Graded Geometry, Batalin-Vilkovisky Formalism and their Applications,''
  Archivum Math.\  {\bf 47}, 143 (2011)
  [arXiv:1105.2680 [math.QA]].

%\bibitem{Roy} D. Roytenberg,  ``On the structure of graded symplectic
%supermanifolds and Courant algebroids''. In {\it Quantization, Poisson Brackets and Beyond}, 
%Theodore Voronov (ed.), Contemp. Math., Vol. 315, Amer. Math. Soc., Providence, RI, (2002)  [arXiv:math/0203110v1 [math.SG]]

\bibitem{Roytenberg:2006qz} D.~Roytenberg, ``AKSZ-BV formalism and Courant
algebroid-induced topological field theories,'' Lett.\ Math.\ Phys.\ \textbf{%
79} (2007) 143 [arXiv:hep-th/0608150].

\bibitem{Schaetz} F. Sch\"atz, ``Invariance of the BFV-complex,'' 
  Pacific J. Math. {\bf 248} (2010), no. 2, 453--474  [arXiv:math/0812.2357]

\bibitem{Sch}
  A.~S.~Schwarz,
  ``Geometry of Batalin-Vilkovisky quantization,''
  Commun.\ Math.\ Phys.\  {\bf 155}, 249 (1993)
  [hep-th/9205088].
 
\bibitem{Signori}  D. Signori, \emph{Sottovarieta coisotrope in teoria di campo e quantizzazione},
Laurea thesis, Milan University, can be downloaded from %http://www.math.uzh.ch/reports/02$\_$05.pdf
http://user.math.uzh.ch/cattaneo/signori.pdf

\bibitem{Stasheff 96} J. D. Stasheff,  ``Homological reduction of
constrained Poisson algebras'', J. Diff. Geom. {\bf 45} (1997)  221-240 [q-alg/9603021].

%\bibitem{Witten} qualcosa sui gauged sigma model

\bibitem{Zuc07} R. Zucchini: ``The Hitchin Model,
Poisson-quasi-Nijenhuis Geometry and Symmetry Reduction'', J.High Energy Phys. {\bf 2007} JHEP10(2007)075 
[arXiv:0706.1289[hep-th]]

\bibitem{Zuc08} R. Zucchini: ``Gauging the Poisson sigma model'', J.High Energy Phys. {\bf 2008} JHEP05(2008)
[arXiv:0801.0655[hep-th]]

\bibitem{Zuc10} R. Zucchini, ``The gauging of BV algebras'', J.Geom.and Phys.{\bf 60} 11 (2010), 1860-1880 [arXiv:1001.0219[hep-th]]
%\bibitem{Zucchini:2008hn}
% R.~Zucchini,
%``The Lie algebroid Poisson sigma model,''
%JHEP {\bf 0812} (2008) 062
%[arXiv:0810.3300 [math-ph]].
\end{thebibliography}
\end{document}